\documentclass[runningheads]{llncs}
\usepackage[utf8]{inputenc}
\usepackage{hyperref}
\usepackage{listings}
\usepackage{subcaption}
\usepackage[export]{adjustbox}
\usepackage{biblatex}
\usepackage{fullpage}
\usepackage{comment}
\title{Hiding Sensitive Information Using PDF Steganography}
\date{}
\author{Ryan Klemm \and 
Bo Chen}
\institute{Department of Computer Science, Michigan Technological University, Michigan, United States
\\
\email{\{rcklemm,bchen\}@mtu.edu}
}
\begin{document}

\maketitle

\begin{abstract}
    The use of steganography to transmit secret data is becoming increasingly common in security products and malware today. Despite being extremely popular, PDF files are not often the focus of steganography research, as most applications utilize digital image, audio, and video files as their cover data. However, the PDF file format is promising for usage in medium-capacity steganography applications. In this paper, we present a novel PDF steganography algorithm based upon least-significant bit insertion into the real-valued operands of PDF stream operators. Where prior research has only considered a small subset of these operators, we take an extensive look at all the possible operators defined in the Adobe PDF standard to evaluate their usability in our steganography algorithm. We also provide a case study which embeds malware into a given cover PDF document. 
\end{abstract}

\section{Introduction}
Steganography is the practice of hiding data inside of a cover media. This cover media is generally a benign file that appears normal in the context of the communication channel that messages are being passed over. The goal of hiding data with steganography is to make it appear as though no hidden information is being transferred at all, so any changes to the cover file must not be obvious to a viewer. The cover media for a steganographic message can be any sort of data, but is most commonly a digital media file such as an image, audio, or video file. This research, however, specifically focuses on hiding sensitive information into the PDF file format, a widely used file format in business, education, and government.

\subsection{PDF Steganography}
A variety of methods have been developed in order to hide data inside of PDF documents. Three of the major categories of PDF steganography 
are described in the following sections. This research will primarily focus on using ``Operator Value'' of the PDF documents for information hiding via steganography. 

\noindent\textbf{Data Outside of View}.
One of the simplest forms of PDF file steganography is placing data in areas of the file in which it will not be rendered when opened by a PDF reader. This could be as simple as placing hidden data within a PDF stream comment or after the \texttt{\%\%EOF} line of the file. More complex strategies include placing hidden data inside an object that is not linked to the main object tree within the PDF -- for example, a page object that is not linked into the document's page catalogue \cite{HiddenPages_2020}. The main drawback to this steganography method is that it can be easily detected and defeated.

\noindent\textbf{Between-Character, Between-Word, Between-Line}.
Similar to other forms of text-file steganography, whitespace is a vector for hiding data within PDF files. One proposed strategy is to switch between Unix (\texttt{\textbackslash n}) and Windows (\texttt{\textbackslash r\textbackslash n}) line breaks within the file to encode one bit per line \cite{OpenPuff_2018}. Other strategies use zero-width characters between characters or words within the text of the PDF, or change the justification of the text to encode data within the size of the whitespace present on the page \cite{Justified_2019}.

\noindent\textbf{Operator Values}.
In order to place text, draw vector graphics, and color the PDF document, PDF files utilize operators to define the layout of a document. Many of these operators take floating-point numerical arguments, which can be changed by very small amounts \cite{PDF_2008}, \cite{PDF_2020}. When these operator arguments (operands) are very slightly changed, almost no visual effects occur within the viewed PDF file \cite{Operators_2008}. Thus, these slight changes can be used as an effective vector for steganography in PDF documents. Only a small subset of the operators defined by the PDF standard have been studied in existing work in steganography, namely \texttt{TJ}, \texttt{Tc}, \texttt{Td}, and \texttt{Tw} \cite{Analysis_2020}.

\subsection{Our Research}
We have proposed a new strategy for embedding sensitive data into PDF documents through imperceptible modifications to the PDF stream operator values. Compared to existing PDF steganography techniques, ours can achieve a higher carrying capacity and embedding rate, while minimizing the visual impacts to the PDF document. Mostly importantly, our work looks through all the possible operators defined in the Adobe PDF standard, identifying all those unique operators which may be utilized to hide sensitive information in a steganographic manner.
\section{Background Information}

\subsection{Structure of a PDF File}

A PDF file can be thought of as a linked tree of objects, each of which is responsible for one piece of the document. For example, a two-page document will have a root object and two page parent objects, each of which is connected to multiple text, image, etc. objects that render the actual visual parts of the page. 

In the file itself, these objects are arranged in a list, with links made by objects referencing the numerical IDs of other objects. Each object consists of an ID, a list of key/value properties that define its type and connections to other objects, and then an optional data stream that holds the actual content such as the text or image data. Below is an example of an object holding a text stream, which displays “ABC”:

\begin{verbatim}
8 0 obj
<<
/Length 72155
>>
stream
BT
/F1 12 Tf 
288 720 Td 
(ABC) Tj
ET
endstream
endobj   
\end{verbatim}

This example shows a PDF object with ID 8, enclosed between the keywords \texttt{obj} and \texttt{endobj}. The stream's properties are stored in between the ‘\texttt{<<}’ and ‘\texttt{>>}’. Following its properties, the object's data stream lies between the keywords \texttt{stream} and \texttt{endstream}. Each line of this stream contains an operator which controls some aspect of the positioning, size, or content of the text displayed \cite{PDF_2008}, \cite{PDF_2020}.

\subsection{PDF Stream Operators}
The method for hiding information inside PDF documents utilizes the stream operators that control the display of the PDF. These operators consist of a string identifier that is preceded by some number of operands, which can either be strings or numerical. For example, the ‘Tf’ operator defines the font and size for the following text:
\begin{equation}
\texttt{/F1 12 Tf}
\end{equation}

\noindent While the number of operands can vary between operators, almost all operators have a defined and consistent number of operands. ‘Tf’ is always preceded by a string that decides the ID of the font to use, and a floating-point precision number that defines the font size. So, the above example denotes the usage of font \#1 at 12-point size. The core of the steganography method comes from the floating-point precision of these operands. In this example, the sequence:

\begin{equation}
\texttt{/F1 12.01 Tf}
\end{equation}

\noindent is also a valid use of the operator, and produces text that is essentially visually identical, since 0.01 points of font size corresponds to negligible height on a printed page. Therefore, replacing sequence (1) with sequence (2) inside of a PDF document has no discernible effect on the visual appearance of the document \cite{PDF_2008}, \cite{PDF_2020}. As will be shown later, sequences (1) and (2) are able to represent different parts of a steganographic hidden message.

\subsubsection{Note on ISO PDF Standards}
The PDF file format was formalized as ISO 32000 in 2008, which defined version 1.7 of PDF documents. Most PDF files in circulation today have versions 1.1-1.7. The steganography method and discussion of the PDF file format in this document are based off of the version 1.7 standard \cite{PDF_2008}. However, in 2020, Adobe \& ISO updated the standard to define PDF version 2.0. While this standard adds some extra features to the PDF file format, it is fully backwards-compatible with older PDF versions:\\

``ISO 32000-1 defines a PDF version matching 1.7. This document [ISO 32000-2] defines a new version of PDF designated 2.0. This document is also suitable for interpretation of files made to conform to any of the previous Adobe PDF specifications 1.0 through 1.7 and ISO 32000-1'' \cite{PDF_2020}.\\

 Therefore, all methods described in this document will continue to be viable for steganography in PDF 2.0 files. There will also be a discussion of if any of the changes made to the standard can be used to expand the steganography method.

\subsection{PDF Management Software}
A variety of software exists to manage PDF files. The primary functionality that is necessary for this paper's steganography method is the ability to compress and decompress PDF streams. In most generated and distributed PDF files, the object streams are encoded with lossless compression in order to reduce the overall size of the PDF file. However, in order to effectively read and write to the streams, they must be decompressed before being operated on by the steganography code. Once the edits are made, the streams can be compressed once again. The following two sections show how to compress and decompress PDF streams in different Linux tools for PDF file management \cite{PDFtk}, \cite{QPDF}. 

\subsubsection{pdftk}
Decompression: \texttt{pdftk <in\_file> output <out\_file> uncompress}\\
Compression: \texttt{pdftk <in\_file> output <out\_file> compress}

\subsubsection{qpdf}
Decompression: \texttt{qpdf <in\_file> --stream-data=uncompress <out\_file>}\\
Compression: \texttt{qpdf <in\_file> --stream-data=compress <out\_file>}

\section{Related Work}
Wang et al.~\cite{Operators_2008} proposed a similar steganographic method to the one that we are proposing. However, they only considered a small subset of the possible operators (\texttt{Tm, MediaBox}). The paper proposed multiple ways to hide data in numerical operands: 1) Adding trailing zeros to all floating-point operands. Thus, to hide the decimal data `5', 5 zeroes would be added to an operand. For example, \texttt{12.1} would change to \texttt{12.100000}. The 5 could then be translated to some binary value. This method is guaranteed to not produce any visual artifacts, since the numerical value of the operands do not change at all. However, it will significantly increase the file size, since each 0 adds 1 character (1 byte) to the file. 2) Adding 2 trailing zeros to a floating-point operand, then appending the numerical value that you wish to encode. For example, to hide the decimal data `123', \texttt{12.1} would change to \texttt{12.100123}. The 123 could then be translated to some binary value. This method slightly improves the issue of increasing the file size. It has the potential to introduce visual artifacts if used with extremely sensitive operators. However, the two trailing zeroes make the percentage magnitude of the numerical change too small to be visually noticeable in almost any case.
Either of these methods could be directly reused on any of the 31 operators that we elaborate in this paper that take floating-point operands.

Sofian et al.~\cite{LSB_AES_2019} proposed a least-significant-bit insertion steganography method on the integer numerical operands to the \texttt{TJ} operator. Considering 98 as a numerical operand, $98_{10} = 01100010_2$. Currently, the least-significant-bit encodes a binary 0. To instead encode a binary 1, the number would be changed to $01100011_2 = 99_{10}$. Thus to encode a 1 instead of a 0, the operand 98 would need to be changed to 99. Since the integer operands in the \texttt{TJ} operator represent thousandths of a unit in text coordinates, a change from 99 to 98 is visually imperceptible \cite{LSB_AES_2019}. Important to note is that encoding 99 as opposed to 98 does not require any extra file space. Therefore, this method can hide data without increasing the size of the cover PDF. In its current state, this steganography method cannot be generally applied to all 31 operators that this paper is concerned with. While a change such as 98 to 99 may be acceptable for the \texttt{TJ} operator, it may produce visual artifacts if used in a different operator.

\section{Our Proposed PDF Steganography Method}
We propose a new PDF stegangraphy method by evaluating all the stream operators defined by the PDF document standard, identifying those which may be leveraged to embed sensitive information. Also, the LSB insertion is the most space efficient way to embed data, while adding trailing digits past the decimal point is the most stable way to avoid visual artifacts after steganography. 

\noindent\textbf{Which Operators can be Used?} To achieve the highest possible capacity with PDF steganography, the exact operands that the method will work on must be defined. The PDF document standard (Version 3.0), published by Adobe, defines 73 different stream operators. Among them, 32 operators (Table~\ref{tbl:operators}) accept numerical operators that can be changed with floating-point precision. These 32 operators may be used for PDF steganography, as they take floating-point precision operands. 
The other 41 operators either have no operands, take string operands, or integer numerical operands that cannot be changed without significantly impacting their visual effect. Therefore, they are very unlikely to be utilized for our purpose.

\begin{table}[tb]
\caption{PDF operators that may be used for PDF steganography}
\centering
\begin{tabular}{c | c | c}
Operator & PDF 2.0 Std. Table & PDF 1.7 Std. Table \\ \hline
c & Table 58 & Table 59 \\ \hline
v & Table 58 & Table 59 \\ \hline
y & Table 58 & Table 59 \\ \hline
l & Table 58 & Table 59 \\ \hline
m & Table 58 & Table 59 \\ \hline
re & Table 58 & Table 59 \\ \hline
cm & Table 56 & Table 57 \\ \hline
i & Table 56 & Table 57 \\ \hline
M & Table 56 & Table 57 \\ \hline
w & Table 56 & Table 57 \\ \hline
G & Table 73 & Table 74 \\ \hline
g & Table 73 & Table 74 \\ \hline
K & Table 73 & Table 74 \\ \hline
k & Table 73 & Table 74 \\ \hline
RG & Table 73 & Table 74 \\ \hline
rg & Table 73 & Table 74 \\ \hline
sc & Table 73 & Table 74 \\ \hline
SC & Table 73 & Table 74 \\ \hline
scn & Table 73 & Table 74 \\ \hline
SCN & Table 73 & Table 74 \\ \hline
Tc & Table 103 & Table 105 \\ \hline
Td & Table 106 & Table 108 \\ \hline
TD & Table 106 & Table 108 \\ \hline
Tf & Table 103 & Table 105 \\ \hline
TL & Table 103 & Table 105 \\ \hline
Tm & Table 106 & Table 108 \\ \hline
Ts & Table 103 & Table 105 \\ \hline
Tw & Table 103 & Table 105 \\ \hline
Tz & Table 103 & Table 105 \\ \hline
TJ & Table 107 & Table 109 \\ \hline
d0 & Table 111 & Table 113 \\ \hline
d1 & Table 111 & Table 113 
\end{tabular}
\label{tbl:operators}
\end{table}

\noindent\textbf{Bit Embedding}. Consider a case where making a change from 98 to 99 may cause visual artifacts. 98 could then instead be represented by 98.0. To do LSB insertion, consider $980_{10} = 0011 1101 0100_2$. This currently encodes a binary 0. To instead encode a 1, flip the LSB: $0011 1101 0101_2 = 981_{10}$. Then, place the decimal point back to its previous location, to reach 98.1. A change from 98.0 to 98.1 may not cause the same visual artifacts as a change from 98 to 99.

Overall, an LSB insertion of the lowest $n$ bits of a number will change the decimal value of that number by a maximum of $2^n - 1$. Perhaps more relevant to PDF operator steganography is the actual percentage change to a numerical value if a certain bit insertion is applied. There is no closed formula for this, as it will vary on a case-by-case basis depending on the bit pattern of the existing numerical value and the bit pattern of the bits that need to be inserted. This percentage change should roughly correspond to how visually noticeable the changes in operand values are. It is crucial that we know the percentage cutoff for each viable steganography operator. This is the purpose of Section 10.

Once the percentage cutoffs are known, the actual steganography algorithm can be carried out:\\
(Let the current operand value be $v$, the maximum allowable percentage change be $p$, the number of bits hidden per operand be $n$, and the next $n$ bits of the hidden message be $S$) 
\begin{enumerate}
    \item Calculate $O$ as the integer formed by the digits of $v$ when the decimal point is removed. Keep track of the number of digits before the decimal point, as the decimal point must be returned to that position after the LSB insertion is carried out.
    
    \item If the least significant $n$ bits of $O$ exactly match $S$, then no change is required. Simply move on to the next relevant operand value. There is a roughly $\frac{1}{2^n}$ chance of this happening for each operand. In this case, $n$ bits of data are hidden without adding any size to the cover file.
    
    \item If the least significant $n$ bits of $O$ do not exactly match $S$, determine the value of $O_S$, formed by inserting $S$ into the least significant $n$ bits of $O$. If $|O_S - O| \le p \cdot O$, then just replace $O$ with $O_S$. 
    
    \item If $|O_S - O| > p \cdot O$, extend $O$ one more digit past the decimal place, so $O_{ext} = 10 \cdot O$. Repeat steps 3 and 4 as many times as necessary until a small enough percentage change has been made. Note that for each application of step 4, one byte is added to the size of the cover file.

    \item Once the final value of $O_S$ is reached, replace the decimal point to form the final operator value.
\end{enumerate}

Using this algorithm, the values of $p$ and $n$ can be easily customized per operator. As long as the recipient knows the value of $n$ used for each operator, they can extract the message. Recommended settings can be found in Table~\ref{tbl:recommended}. We have further justified the choices made in the `Percentage per Operand' field of Table~\ref{tbl:recommended} in Appendix~\ref{app:evaluation}.

\begin{table}[tb]
\caption{Minimum percentage change necessary for usable operators}
\centering
\begin{tabular}{c | c | c | c}
Operator & Num (Usable) Operands & Percentage per Operand & Reliability \\ \hline
c & 6 & 1 & good \\ \hline
v & 4 & 1 & good \\ \hline
y & 4 & 1 & good \\ \hline
l & 2 & 0.05 & good \\ \hline
m & 2 & 0.05 & good \\ \hline
re & 4 & 0.2 & good \\ \hline
cm & 6 & 0.1, 0.05 & good \\ \hline
i & 1 & -- & good \\ \hline
M & 1 & -- & good \\ \hline
w & 1 & 1 & good \\ \hline
G & 1 & 5 & good \\ \hline
g & 1 & 5 & good \\ \hline
K & 4 & 5 & good \\ \hline
k & 4 & 5 & good \\ \hline
RG & 3 & 5 & good \\ \hline
rg & 3 & 5 & good \\ \hline
sc & 1, 3, 4 & 5 & good \\ \hline
SC & 1, 3, 4 & 5 & good \\ \hline
scn & 1, 3, 4 & 5 & good \\ \hline
SCN & 1, 3, 4 & 5 & good \\ \hline
Tc & 1 & 1 & low \\ \hline
Td & 2 & 2 & good \\ \hline
TD & 2 & 2 & good \\ \hline
Tf & 1 & 0.5 & good \\ \hline
TL & 1 & 2 & good \\ \hline
Tm & 6 & 5-10, 2 & good \\ \hline
Ts & 1 & 5-10 & good \\ \hline
Tw & 1 & 1 & low \\ \hline
Tz & 1 & 0.5 & good \\ \hline
TJ & 0$+$ & 15 & good \\ \hline
d0 & 1 & 1 & good \\ \hline
d1 & 5 & 1, 1-2 & good
\end{tabular}
\label{tbl:recommended}
\end{table}

\noindent\textbf{Bit Extraction}. 
Bit extraction is a significantly simpler process than bit embedding. Given an operator value $v$ and the number of embedded bits $n$, the extraction is straightforward:
\begin{enumerate}
    \item Form the mask $m$ as the binary string of $n$ 1's. 
    \item Calculate $O$ as the integer formed by the digits of $v$ with the decimal point removed.
    \item Calculate the next $n$ binary digits of the hidden message by extracting the least significant $n$ bits of $O$
\end{enumerate}

\noindent\textbf{Regex for Detecting Operators in a PDF}. The most effective way of locating the operator sequences of a PDF document within code is through the use of regex masks. The regex mask for all operators except for \texttt{TJ} is as follows:\\
\lstinline|(?:[\d\.\-]+\s+){a, b}op[\[\s]| \\
Where $a$ and $b$ are the minimum and maximum number of operands for the given operator with code $op$.

The regex mask for operator \texttt{TJ} is as follows:\\
\lstinline|\[.+?\]\s*?TJ|\\

The above two regex masks pull out the full operator sequence from the PDF file streams. To grab the individual numerical operands, a further regex operation is necessary. The regex mask for extracting operands from all operators except for \texttt{TJ} is as follows:\\
\lstinline|[\d\.\-]+|

\noindent The regex mask for extracting operands from operator \texttt{TJ} is as follows:\\
\lstinline|[\d\.\-]+(?![^\(]*\))(?![^\<]*\>)|


\section{Case Study: Leveraging PDF steganography to Hide Malware}

To show how to use the proposed PDF steganography method for practical purposes, we have implemented the proposed design in Python (the code is available in~\cite{pdfsteg}) and performed a case study in which we embed a malware sample into a given cover PDF. We start by obtaining our cover PDF and malware sample:\\

\noindent Cover PDF -- Adobe PDF Standard Specification, Version 1.7 -- Source: \url{https://opensource.adobe.com/dc-acrobat-sdk-docs/pdfstandards/PDF32000_2008.pdf}\\
Malware Sample -- Petya Ransomware Trojan (Windows) -- Source: \url{https://github.com/ytisf/theZoo/tree/master/malware/Binaries/Trojan.Ransom.Petya}\\

\noindent Information on the malware sample file is shown in the below image:

\includegraphics[scale=0.5]{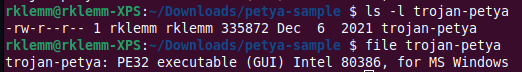}\\

Before we can operate on the PDF document, we must decompress the streams using one of the methods discussed in Section 2.4:\\
\texttt{\$ qpdf PDF32000\_2008.pdf --stream-data=uncompress cover.pdf}\\
This yields the file \texttt{cover.pdf}, which has all object stream data in cleartext.\\

Before attempting to embed the data, it is important to check the actual capacity of the cover PDF document. With our steganography Python script, that is done with the following command:\\
\texttt{\$ python3 full\_pdf\_steg.py stat cover.pdf}\\
Which gives the results shown:

\includegraphics[scale=0.5, trim={0 0 11in 0.6cm}, clip]{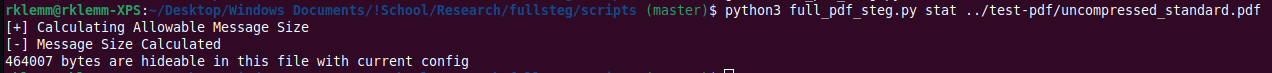}\\

Our malware sample has a size of 335,872 bytes, which is clearly smaller than the 464,007 byte carrying capacity of the cover PDF. Thus, we can continue on with the process. The command to embed the hidden data within the PDF is the following:\\
\texttt{\$ python3 full\_pdf\_steg.py embed cover.pdf steg.pdf trojan-petya}\\
The results of this command are shown below:\\

\includegraphics[scale=0.5, trim={0 0 13.170in 1.25cm}, clip]{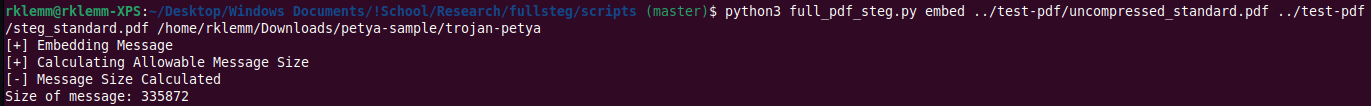}

\vspace{-0.075cm}
\includegraphics[scale=0.5]{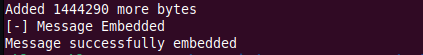}

After the steganography algorithm finishes processing the PDF file, the changes made in the document are almost imperceptible. To show an example, a page before (left) and after (right) data hiding is shown below:

\includegraphics[trim={0.5cm 0.5cm 0.5cm 0.5cm}, clip, scale=0.25]{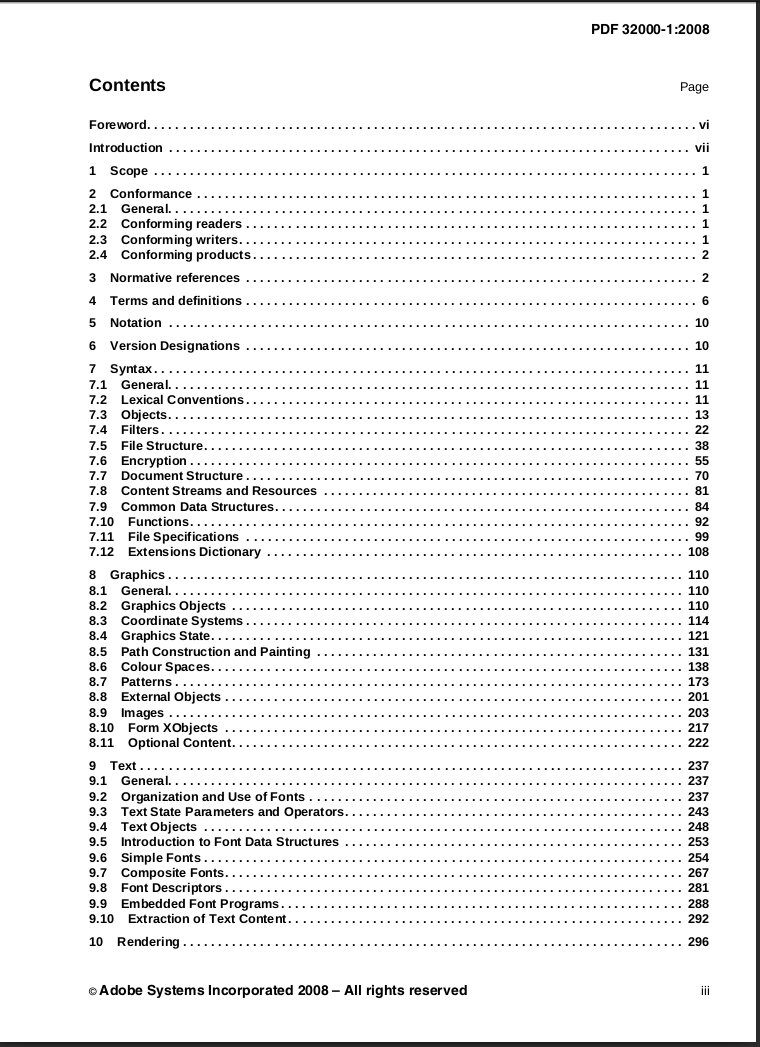}
\includegraphics[trim={0.5cm 0.5cm 0.5cm 0.5cm}, clip, scale=0.25]{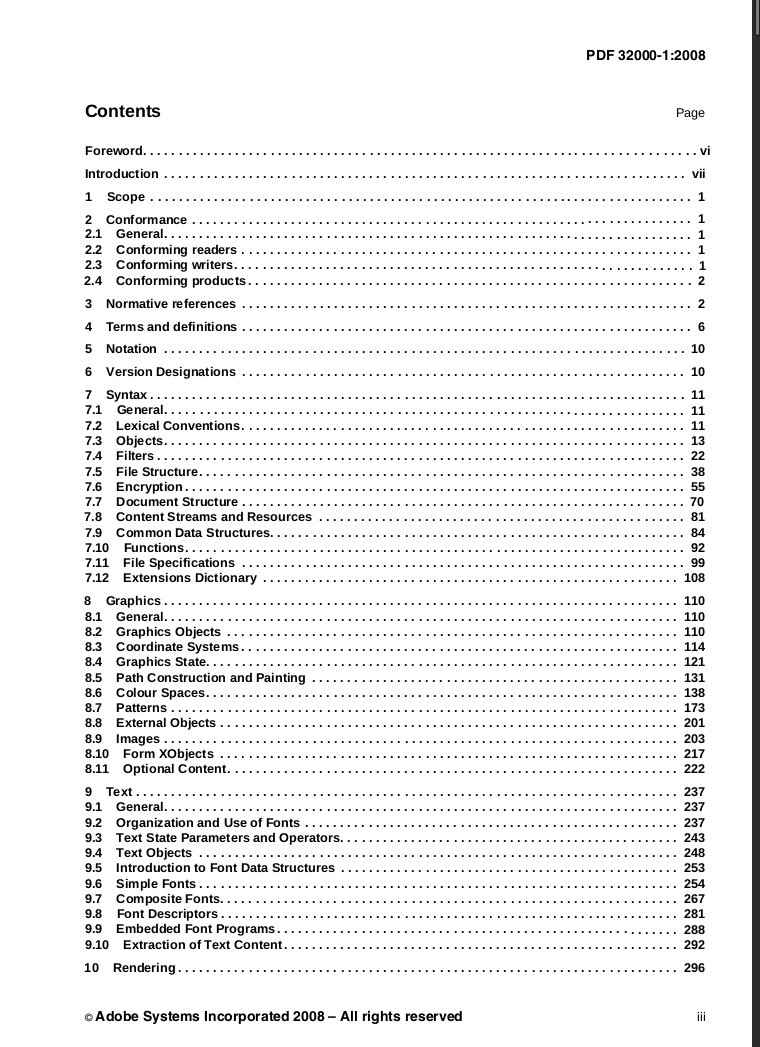}

The above command creates the file \texttt{steg.pdf}, which has the malware steganographically encoded into it. To check that our method is reliable, we can now extract the malware sample back out of \texttt{steg.pdf}. This is performed with the following command:\\
\texttt{\$ python3 full\_pdf\_steg.py extract steg.pdf trojan-petya-out}\\
The results of this command are shown below:\\

\includegraphics[scale=0.5, trim={0 0 15in 1.325cm}, clip]{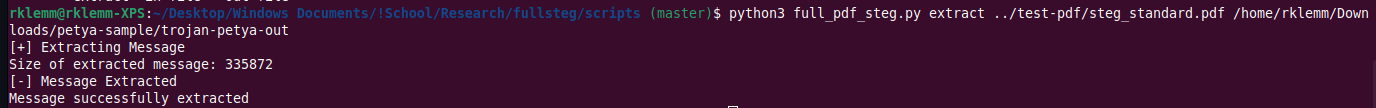}\\

We see that the size of the extracted file matches the size of our original malware sample. To further confirm that the embedded and extracted malware samples are the same, we could run an \texttt{md5sum} command on both files and confirm that the hashes are equal.\\

The \texttt{steg.pdf} file still has uncompressed object streams, and therefore is a much larger file than the original \texttt{PDF32000\_2008.pdf} file. To recover disk space, we can (lossless) re-compress the PDF with the following command, as discussed in Section 2.4:\\
\texttt{\$ qpdf steg.pdf --stream-data=compress compressed\_steg.pdf}\\

The sizes of all relevant files are shown in the table below:\\
\begin{tabular}{c | c | c}
File Name & Compressed Streams & File Size (bytes) \\ \hline
\texttt{PDF32000\_2008.pdf} & Y & 22,491,828\\ \hline
\texttt{cover.pdf} & N & 30,840,706 \\ \hline
\texttt{steg.pdf} & N & 32,284,996 \\ \hline
\texttt{compressed\_steg.pdf} & Y & 22,909,867 \\ \hline
\texttt{trojan-petya} & N/A & 335,872 \\
\end{tabular}\\

\noindent The following calculations summarize the results of this case study.\\

\noindent Steg message capacity = 464,007 bytes\\

\noindent Net Increase in PDF size (decompressed streams) = 1,444,290 bytes\\

\noindent Net Increase in PDF size (compressed streams) = 418,039 bytes\\

\noindent Rate (steg message capacity / compressed cover size) = 2.06\% \\

\noindent Rate (steg message size / decompressed size increase) = 23.25\% \\

\noindent Rate (steg message size / compressed size increase) = 80.34\% \\

\noindent Rate (decompressed size increase / decompressed cover size) = 4.68\% \\

\noindent Rate (compressed size increase / compressed cover size) = 1.86\% \\

Overall, this sample run of the PDF steganography method successfully embedded a 335 KB malware sample into a 22.5 MB PDF file. The resulting PDF file has a size of 22.9 MB, which is an net size increase of less than 2\%. The steganographic encoding very efficiently uses the additional size that it adds to the PDF file.

\section{Conclusion}
By utilizing an adaptable strategy of least-significant bit insertion on string representations of floating-point operands to perform imperceptible changes to PDF stream operators, this paper presents a novel PDF steganography method. Our method has a higher carrying capacity than previous PDF operator-based methods due to the use of all viable operators found within a PDF file. Our case study of embedding malware into a given PDF cover file justifies effectiveness of our proposed approach. 



\noindent\textbf{Acknowledgments.} This work was supported by US National Science Foundation under grant number CNS-1928349, CNS-2225424, and 2043022-DGE.

\printbibliography


\appendix

\clearpage
\section{Evaluating All Possible Operators for PDF Steganography}
\label{app:evaluation}
Note that \textit{the quoted descriptions of operators in below sections are pulled directly from the PDF standard}.

\subsection{Graphics Drawing Operators}
The following operators control graphics that are drawn onto a PDF document. They generally take operands in units of the coordinate space of the PDF document.

\texttt{c, v, y} operators draw cubic Bézier curves based on the coordinate parameters given. The difference between the variants determines which points are used as endpoints, and which points are used as control points.

\texttt{l, m} operators move the `current point' of the PDF graphics. \texttt{m} skips directly to the new point, while \texttt{l} draws a straight line from the previous point to the new point. \texttt{m} is often used to start a sequence of graphics operations.

\texttt{re} operator draws a rectangle.

\texttt{cm} operator specifies the matrix transform to apply to all following graphics.

\texttt{i, M, w} each use one operand to set a single graphics parameter.

Overall, the study of using this class of operators for steganography shows that graphics coordinates should likely not be changed by integral amounts, which impacts operators \texttt{c, v, y, l, m, re}.

\subsubsection{c Operator}
Takes 6 floating-point operands: \texttt{x1 y1 x2 y2 x3 y3}.

Table 58: ``Append a cubic Bézier curve to the current path. The curve
shall extend from the current point to the point (x3 , y3 ), using
(x1 , y1 ) and (x2 , y2 ) as the Bézier control points (see 8.5.2.2,
"Cubic Bézier Curves"). The new current point shall be
(x3 , y3 ).'' \cite{PDF_2020}

Figure 1 shows test runs of the \texttt{c} operator, with differing percentages of change in the coordinates. It appears that a percentage change of 0.1\% in the operator values causes the least amount of visible change. This makes sense, since it corresponds to non-integer change in the PDF coordinate space. For any operators that require the specification of coordinate points, it appears that taking changes (at least) one place past the decimal point is the safest option.

Cutoff percentage will be equivalent to \texttt{v} and \texttt{y} operators, since this is just a different ordering of the same operation.

\subsubsection{v Operator}
Takes 4 floating-point operands: \texttt{x2 y2 x3 y3}

Table 58: ``Append a cubic Bézier curve to the current path. The curve
shall extend from the current point to the point (x3 , y3 ), using
the current point and (x2 , y2 ) as the Bézier control points (see
8.5.2.2, "Cubic Bézier Curves"). The new current point shall
be (x3 , y3 ).'' \cite{PDF_2020}

Cutoff percentage will be equivalent to \texttt{c} and \texttt{y} operators, since this is just a different ordering of the same operation.

\subsubsection{y Operator}
Takes 4 floating-point operands: \texttt{x1 y1 x3 y3}

Table 58: ``Append a cubic Bézier curve to the current path. The curve
shall extend from the current point to the point (x3 , y3 ), using
(x1 , y1 ) and (x3 , y3 ) as the Bézier control points (see 8.5.2.2,
"Cubic Bézier Curves"). The new current point shall be
(x3 , y3 ).'' \cite{PDF_2020}

Cutoff percentage will be equivalent to \texttt{c} and \texttt{v} operators, since this is just a different ordering of the same operation.

\clearpage

\begin{figure}[h]
\centering
\begin{subfigure}{0.45\textwidth}
\centering
\includegraphics[width=0.9\linewidth, height=4cm]{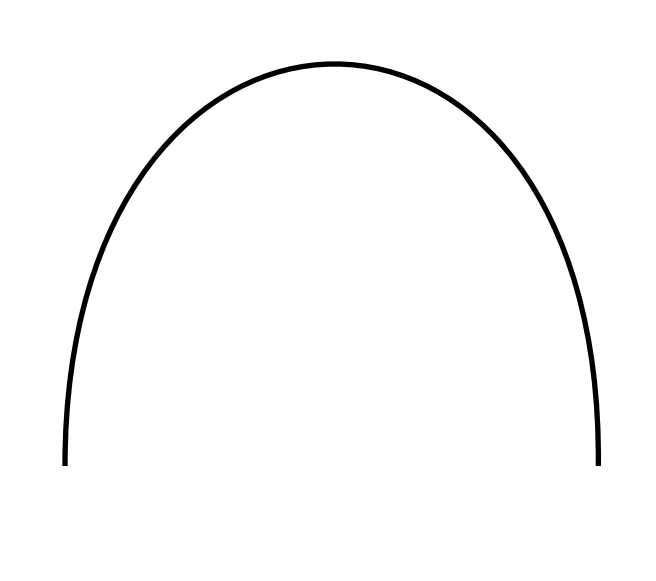}
\caption{\texttt{300 400 400 400 400 300 c}}
\label{fig:c_subim1}
\end{subfigure}
\begin{subfigure}{0.45\textwidth}
\centering
\includegraphics[width=0.9\linewidth, height=4cm]{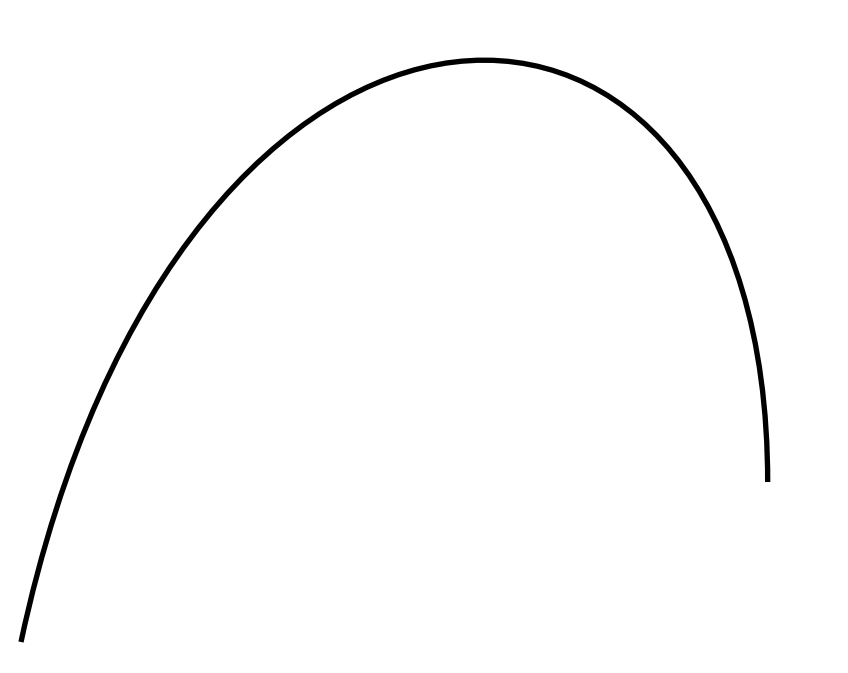}
\caption{\texttt{330 440 440 440 440 330 c}}
\label{fig:c_subim2}
\end{subfigure}

\begin{subfigure}{0.45\textwidth}
\centering
\includegraphics[width=0.9\linewidth, height=4cm]{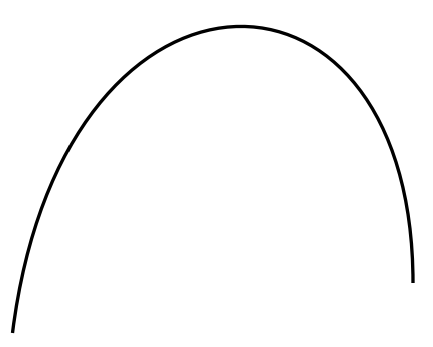}
\caption{\texttt{315 420 420 420 420 315 c}}
\label{fig:c_subim3}
\end{subfigure}
\begin{subfigure}{0.45\textwidth}
\centering
\includegraphics[width=0.9\linewidth, height=4cm]{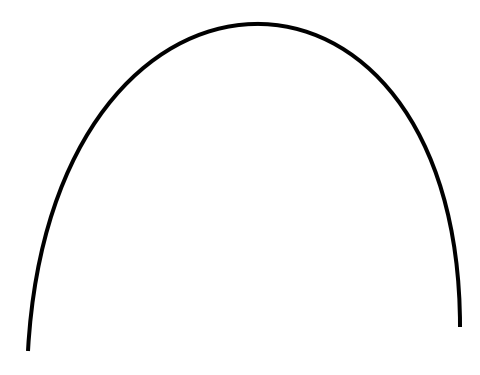}
\caption{\texttt{306 408 408 408 408 306 c}}
\label{fig:c_subim4}
\end{subfigure}

\begin{subfigure}{0.45\textwidth}
\centering
\includegraphics[width=0.9\linewidth, height=4cm]{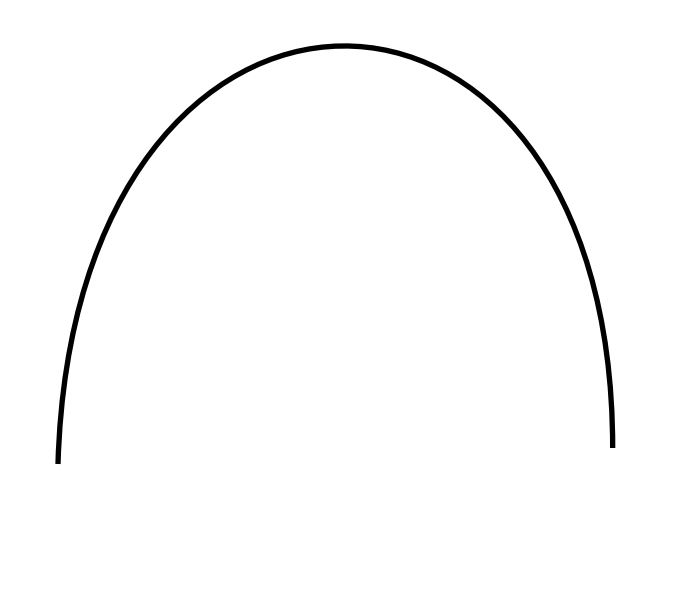}
\caption{\texttt{303 404 404 404 404 303 c}}
\label{fig:c_subim5}
\end{subfigure}
\begin{subfigure}{0.45\textwidth}
\centering
\includegraphics[width=0.9\linewidth, height=4cm]{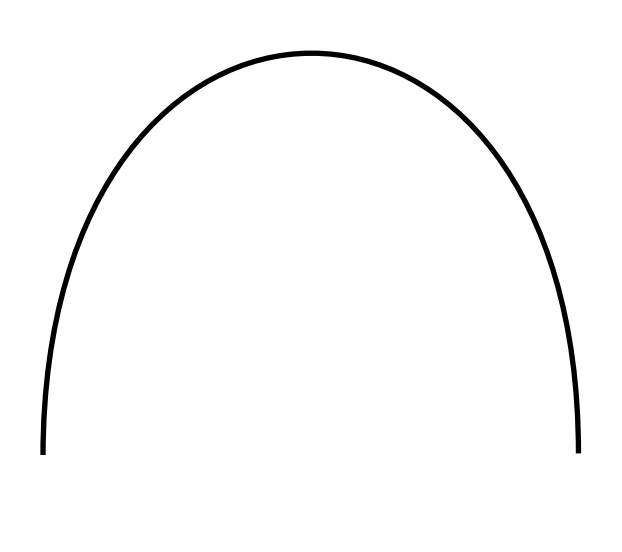}
\caption{\texttt{300.3 400.4 400.4 400.4 400.4 300.3 c}}
\label{fig:c_subim6}
\end{subfigure}
\caption{c Operator - control, 10\%, 5\%, 2\% 1\%, 0.1\% (1\%)}
\label{fig:c_op}
\end{figure}

\subsubsection{l Operator}
Takes 2 floating-point operands: \texttt{x y}

Table 58: ``Append a straight line segment from the current point to the
point (x, y). The new current point shall be (x, y).''

\subsubsection{m Operator}
Takes 2 floating-point operands: \texttt{x y}

Table 58: ``Begin a new subpath by moving the current point to
coordinates (x, y), omitting any connecting line segment. If
the previous path construction operator in the current path
was also m, the new m overrides it; no vestige of the
previous m operation remains in the path.'' \cite{PDF_2020}

The \texttt{m} operator places a point for PDF graphics control. Thus, it will impact the starting point of the next graphics operator. Since it acts on coordinates in graphics space, it's cutoff shall be similar to \texttt{c}, \texttt{l}, \texttt{re} operators. We will place the cutoff at 0.05\%.


\begin{figure}[h]
\centering
\begin{subfigure}{0.45\textwidth}
\centering
\includegraphics[width=0.9\linewidth, height=4cm]{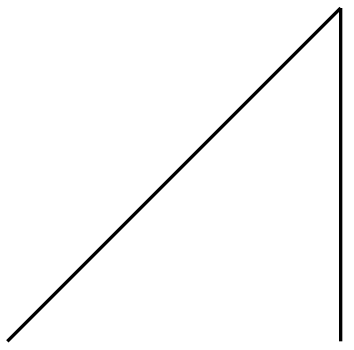}
\caption{\texttt{500 600 l}}
\label{fig:l_subim1}
\end{subfigure}
\begin{subfigure}{0.45\textwidth}
\centering
\includegraphics[width=0.9\linewidth, height=4cm]{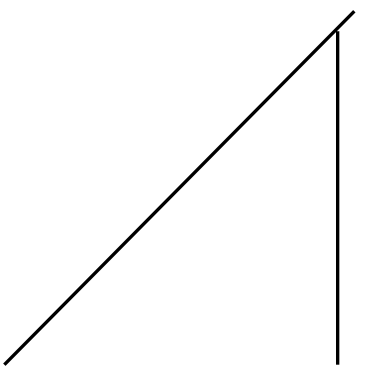}
\caption{\texttt{505 606 l}}
\label{fig:l_subim2}
\end{subfigure}

\begin{subfigure}{0.45\textwidth}
\centering
\includegraphics[width=0.9\linewidth, height=4cm]{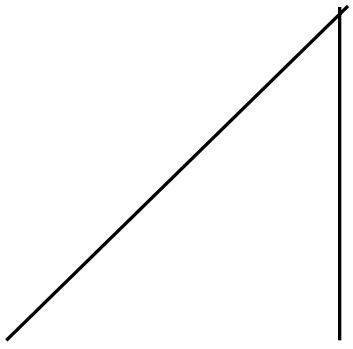}
\caption{\texttt{502.5 603 l}}
\label{fig:l_subim3}
\end{subfigure}
\begin{subfigure}{0.45\textwidth}
\centering
\includegraphics[width=0.9\linewidth, height=4cm]{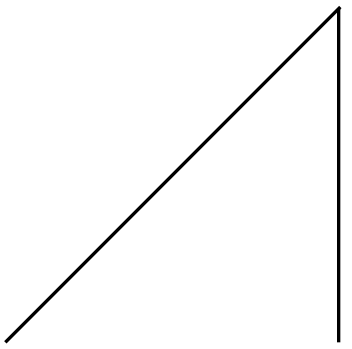}
\caption{\texttt{500.5 600.6 l}}
\label{fig:l_subim4}
\end{subfigure}

\begin{subfigure}{0.45\textwidth}
\centering
\includegraphics[width=0.9\linewidth, height=4cm]{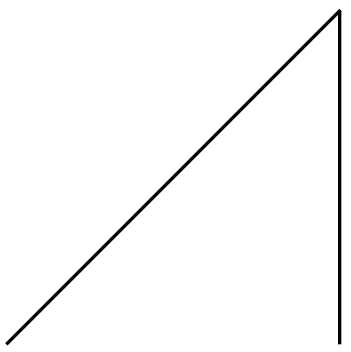}
\caption{\texttt{500.25 600.3 l}}
\label{fig:l_subim5}
\end{subfigure}
\begin{subfigure}{0.45\textwidth}
\centering
\includegraphics[width=0.9\linewidth, height=4cm]{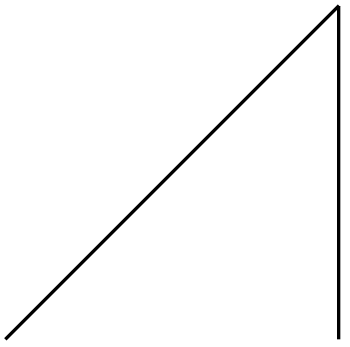}
\caption{\texttt{500.1 600.12 l}}
\label{fig:l_subim6}
\end{subfigure}
\caption{l Operator - control, 1\%, 0.5\%, 0.1\% 0.05\%, 0.02\% (0.05\%)}
\label{fig:l_op}
\end{figure}

\subsubsection{re Operator}
Takes 4 floating-point operand: \texttt{x y width height}

Table 58: ``Append a rectangle to the current path as a complete
subpath, with lower-left corner (x, y) and dimensions width
and height in user space.'' \cite{PDF_2020}

This operator is equivalent to drawing 4 lines of a rectangle with the \texttt{l} operator.

\begin{figure}[h]
\centering
\begin{subfigure}{0.45\textwidth}
\centering
\includegraphics[width=0.9\linewidth, height=4cm]{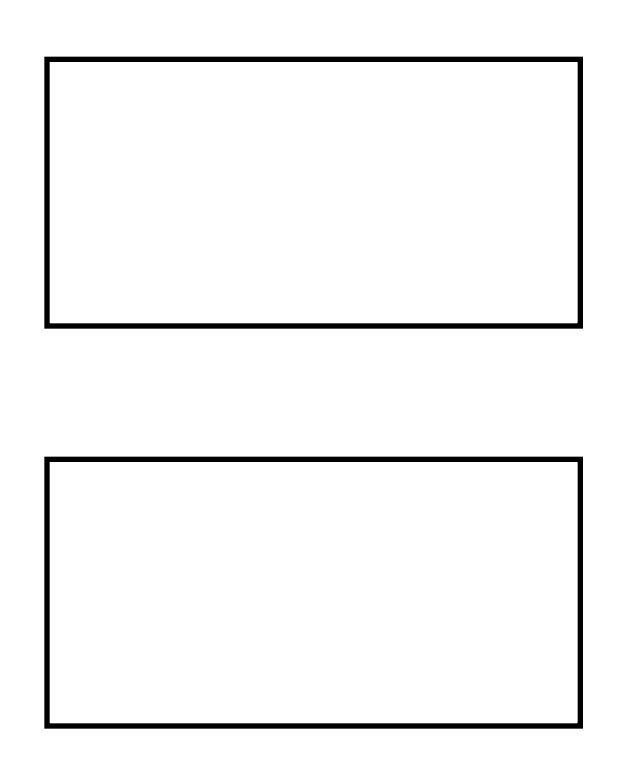}
\caption{\texttt{300 675 100 50 re\\ 300 600 100 50 re}}
\label{fig:re_subim1}
\end{subfigure}
\begin{subfigure}{0.45\textwidth}
\centering
\includegraphics[width=0.9\linewidth, height=4cm]{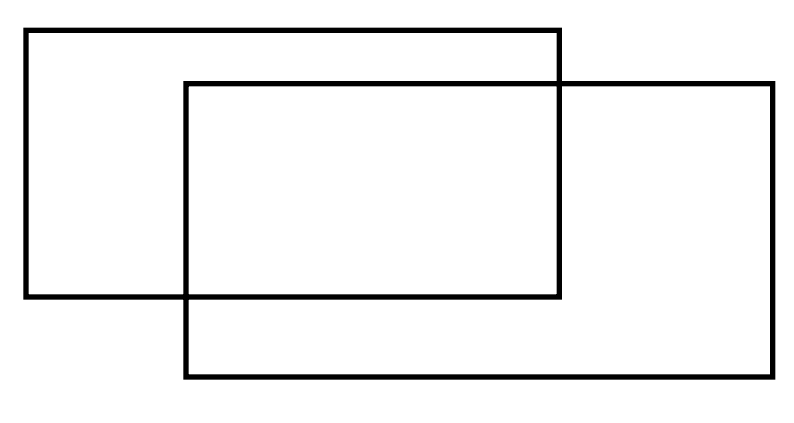}
\caption{\texttt{300 675 100 50 re\\ 330 660 110 55 re}}
\label{fig:re_subim2}
\end{subfigure}

\begin{subfigure}{0.45\textwidth}
\centering
\includegraphics[width=0.9\linewidth, height=4cm]{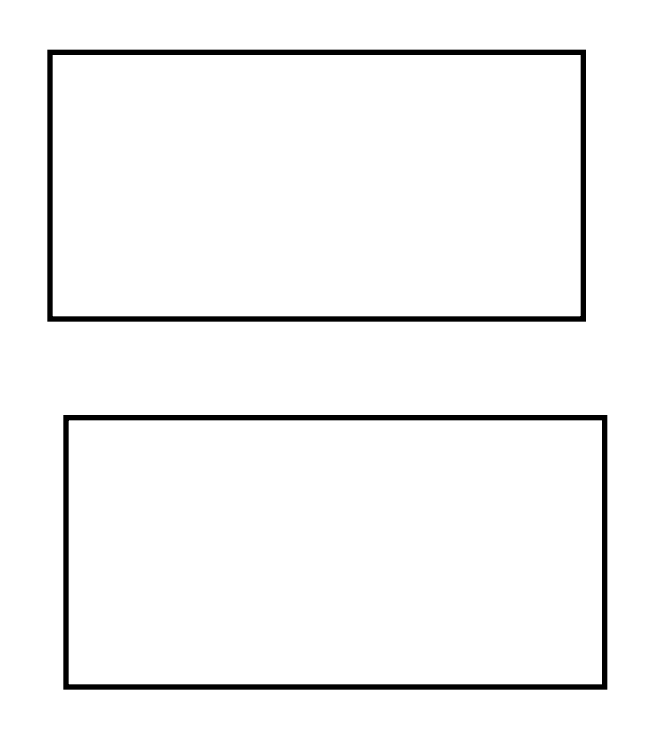}
\caption{\texttt{300 675 100 50 re\\ 303 606 101 50.5 re}}
\label{fig:re_subim3}
\end{subfigure}
\begin{subfigure}{0.45\textwidth}
\centering
\includegraphics[width=0.9\linewidth, height=4cm]{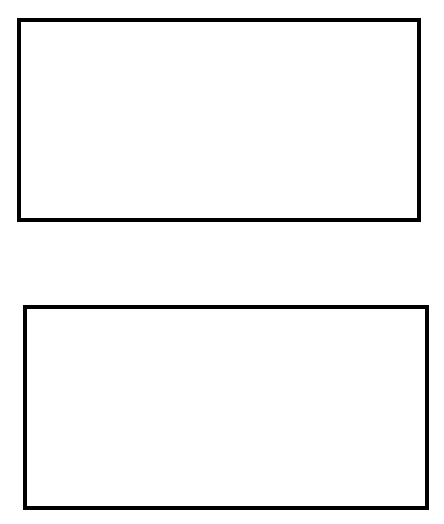}
\caption{\texttt{300 675 100 50 re\\ 301.5 603 100.5 50.25 re}}
\label{fig:re_subim4}
\end{subfigure}

\begin{subfigure}{0.45\textwidth}
\centering
\includegraphics[width=0.9\linewidth, height=4cm]{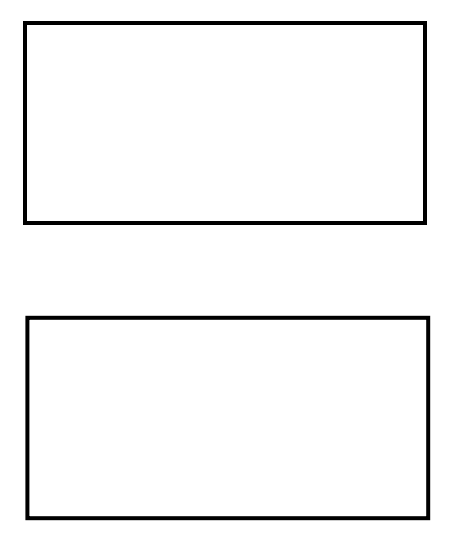}
\caption{\texttt{300 675 100 50 re\\ 300.6 601.2 100.2 50.1 re}}
\label{fig:re_subim5}
\end{subfigure}
\begin{subfigure}{0.45\textwidth}
\centering
\includegraphics[width=0.9\linewidth, height=4cm]{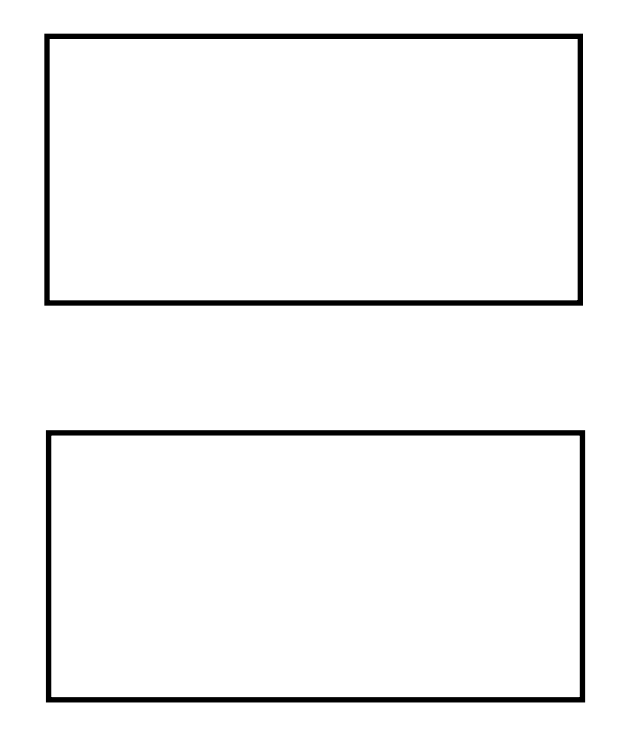}
\caption{\texttt{300 675 100 50 re\\ 300.3 600.6 100.1 50.05 re}}
\label{fig:re_subim6}
\end{subfigure}

\caption{re Operator - control, 10\%, 1\%, 0.5\%, 0.2\% 0.1\% (0.2\%)}
\label{fig:re_op}
\end{figure}

\clearpage

\subsubsection{cm Operator}
Takes 6 floating-point operands: \texttt{a b c d e f}.

Table 56: ``Modify the current transformation matrix (CTM) by concatenating
the specified matrix (see 8.3.2, "Coordinate Spaces"). Although the
operands specify a matrix, they shall be written as six separate
numbers, not as an array.'' \cite{PDF_2020}

Operands \texttt{e} and \texttt{f} simply shift the graphics space simply shift the graphics space by \texttt{e} units in the x direction and \texttt{f} units in the y direction. This makes those operands very similar in effect and units to the \texttt{m} operator, so we will place the cutoff for operands \texttt{e} and \texttt{f} at 0.1\%. 

Operands \texttt{a, b, c, d} are more complex, as they apply a matrix transformation to the graphics space. Just modifying \texttt{a} and \texttt{d} will create stretching or compressing in the horizontal and vertical directions. Just modifying \texttt{b} and \texttt{c} will create skew of the vertical and horizontal axes. Modifying all 4 will cause a combination of both effects, including rotation of the coordinate space.

\clearpage

\begin{figure}[h]
\centering
\begin{subfigure}{0.45\textwidth}
\centering
\includegraphics[width=0.9\linewidth, height=4cm]{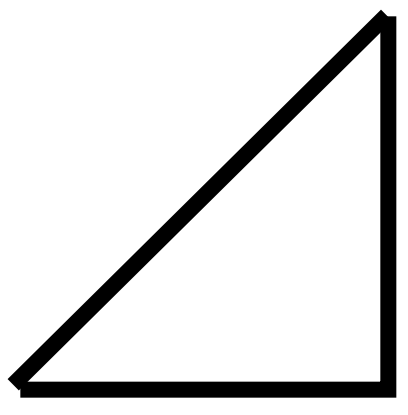}
\caption{\texttt{0.5 0.866 -0.866 0.5 0 0 cm}}
\label{fig:cm_subim1}
\end{subfigure}
\begin{subfigure}{0.45\textwidth}
\centering
\includegraphics[width=0.9\linewidth, height=4cm]{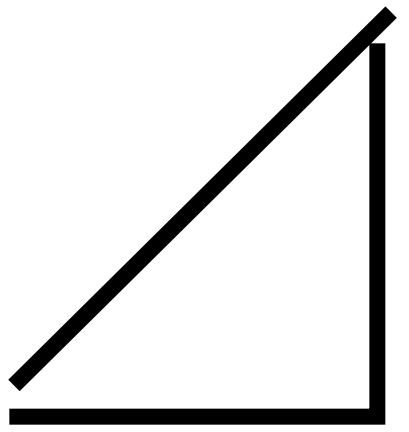}
\caption{\texttt{0.505 0.874 -0.874 0.505 0 0 cm}}
\label{fig:cm_subim2}
\end{subfigure}

\begin{subfigure}{0.45\textwidth}
\centering
\includegraphics[width=0.9\linewidth, height=4cm]{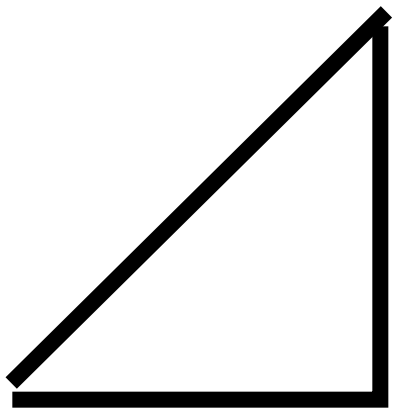}
\caption{\texttt{0.5025 0.8703 -0.8703 0.5025 0 0 cm}}
\label{fig:cm_subim3}
\end{subfigure}
\begin{subfigure}{0.45\textwidth}
\centering
\includegraphics[width=0.9\linewidth, height=4cm]{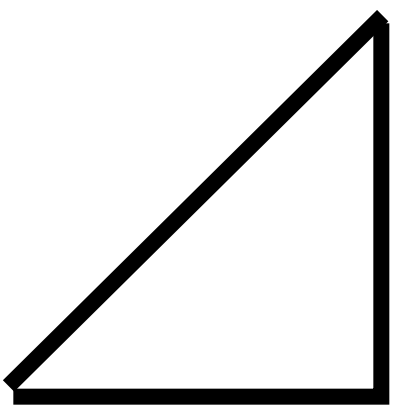}
\caption{\texttt{0.501 0.8677 -0.8677 0.501 0 0 cm}}
\label{fig:cm_subim4}
\end{subfigure}

\begin{subfigure}{0.45\textwidth}
\centering
\includegraphics[width=0.9\linewidth, height=4cm]{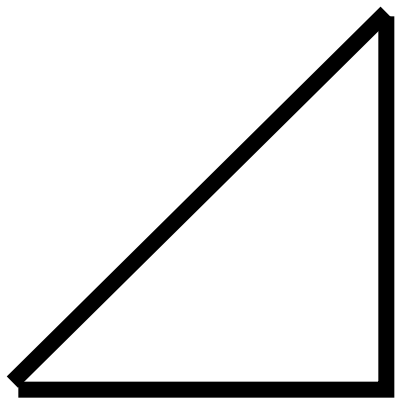}
\caption{\texttt{0.5005 0.86687 -0.86687 0.5005 0 0 cm}}
\label{fig:cm_subim5}
\end{subfigure}
\begin{subfigure}{0.45\textwidth}
\centering
\includegraphics[width=0.9\linewidth, height=4cm]{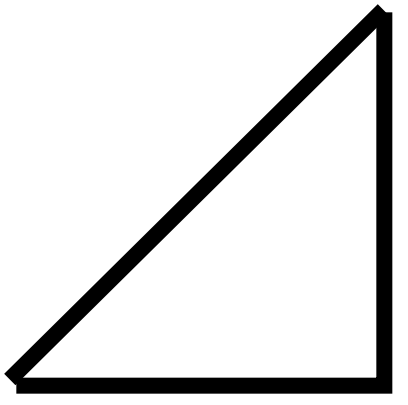}
\caption{\texttt{0.50025 0.86643 -0.86643 0.50025 0 0 cm}}
\label{fig:cm_subim6}
\end{subfigure}

\caption{cm Operator - control, 1\%, 0.5\%, 0.2\% 0.1\%, 0.05\% (0.1\%)}
\label{fig:cm_op}
\end{figure}

\clearpage

\subsubsection{i Operator (?)}
Takes 1 floating-point operand: \texttt{flatness}

Table 56: ``Set the flatness tolerance in the graphics state (see 10.6.2,
"Flatness Tolerance"). flatness is a number in the range 0 to 100; a
value of 0 shall specify the output device’s default flatness
tolerance.'' \cite{PDF_2020}

``The flatness tolerance controls the maximum permitted distance in device pixels between the mathematically
correct path and an approximation constructed from straight line segments, as shown in Figure 54. Flatness
may be specified as the operand of the i operator (see Table 57) or as the value of the FL entry in a graphics
state parameter dictionary (see Table 58). It shall be a positive number.''

\clearpage

\subsubsection{M Operator (?)}
Takes 1 floating-point operand: \texttt{miterLimit}

Table 56: ``Set the miter limit in the graphics state (see 8.4.3.5, "Miter Limit").'' \cite{PDF_2020}

The formula given in the standard for miter limit is given as: $M = \frac{1}{\sin(\frac{j}{2})}$, where $j$ is the angle between two meeting lines. If the calculated value for two arbitrary meeting lines is greater than the provided miter limit, then the point where the lines meet is turned into a bevel.

From the above formula, we can calculate the angle from the miter limit as follows: $j = 2 \cdot \arcsin(\frac{1}{M})$ 

Section 8.4.3.5: ``A miter limit of 1.414 converts miters to bevels for j less than 90 degrees, a limit of 2.0 converts them for j less than 60 degrees, and a limit of 10.0 converts them for j less than approximately 11.5 degrees.'' \cite{PDF_2020}

The following table looks at the percentage change in miter limit vs percentage change in the miter cutoff angle, for initial miter limits of 1.414, 2, and 10.

\begin{tabular}{c | c | c | c}
Miter Limit & Angle & \% change Miter Limit & \% change Angle \\ \hline
1.414 & 90.02 & 0 & 0 \\
1.555 & 80.04 & 10 & 11 \\
1.485 & 84.66 & 5 & 6 \\
1.442 & 87.81 & 2 & 2.5 \\
1.428 & 88.9 & 1 & 1.25 \\ \hline \\ \hline
2 & 60 & 0 & 0 \\
2.2 & 54.07 & 10 & 10 \\
2.1 & 56.87 & 5 & 5.2 \\
2.04 & 58.71 & 2 & 2.2 \\
2.02 & 59.35 & 1 & 1.1 \\ \hline \\ \hline
10.0 & 11.48 & 0 & 0 \\
11.0 & 10.43 & 10 & 9.1 \\
10.5 & 10.93 & 5 & 4.79 \\
10.2 & 11.25 & 2 & 2 \\
10.1 & 11.36 & 1 & 1.05
\end{tabular}\\

It can be seen from the table that the percentage change in the miter limit correlates roughly linearly with the change in the cutoff angle between miters and bevels. This is likely due to the fact that the sine function can be accurately modelled as a linear function for small changes. Thus, by determining how much we can allow our angle cutoff to change, we can apply the same cutoff to the miter limit operand.

\subsubsection{w Operator}
Takes 1 floating-point operand: \texttt{lineWidth}

Table 56: ``Set the line width in the graphics state (see 8.4.3.2, "Line Width").'' \cite{PDF_2020}

\begin{figure}[h]
\centering
\begin{subfigure}{0.45\textwidth}
\centering
\includegraphics[width=0.9\linewidth, height=2cm]{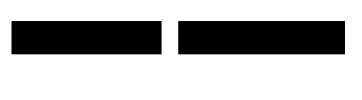}
\caption{\texttt{10 w \\ 10 w}}
\label{fig:w_subim1}
\end{subfigure}
\begin{subfigure}{0.45\textwidth}
\centering
\includegraphics[width=0.9\linewidth, height=2cm]{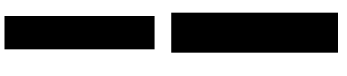}
\caption{\texttt{10 w \\ 12 w}}
\label{fig:w_subim2}
\end{subfigure}

\begin{subfigure}{0.45\textwidth}
\centering
\includegraphics[width=0.9\linewidth, height=2cm]{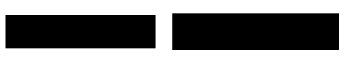}
\caption{\texttt{10 w \\ 11 w}}
\label{fig:w_subim3}
\end{subfigure}
\begin{subfigure}{0.45\textwidth}
\centering
\includegraphics[width=0.9\linewidth, height=2cm]{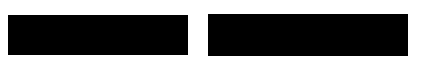}
\caption{\texttt{10 w \\ 10.5 w}}
\label{fig:w_subim4}
\end{subfigure}

\begin{subfigure}{0.45\textwidth}
\centering
\includegraphics[width=0.9\linewidth, height=2cm]{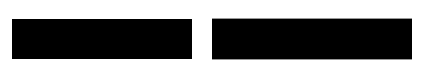}
\caption{\texttt{10 w \\ 10.2 w}}
\label{fig:w_subim5}
\end{subfigure}
\begin{subfigure}{0.45\textwidth}
\centering
\includegraphics[width=0.9\linewidth, height=2cm]{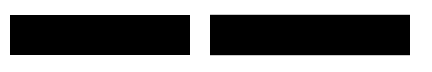}
\caption{\texttt{10 w \\ 10.1 w}}
\label{fig:w_subim6}
\end{subfigure}

\caption{w Operator - control, 20\%, 10\%, 5\%, 2\% 1\% (1\%)}
\label{fig:w_op}
\end{figure}

\subsection{Color Control Operators}
The following operators control the color of text and graphics in a PDF document. They generally take operands in units of color intensity in some color space. Operators specified by capital letters control colors for `stroking' operations, which draw the outline of shapes. Operators specified by lowercase letters control the colors for `fill' operations, which draw the inside of shapes. 

Overall, we can assign a uniform percentage cutoff of 5\% to all the color operators. A 5\% change in color intensity is very difficult to see visually. This lines up with existing methods for image steganography that handle changes in color.

\subsubsection{G Operator}
Takes 1 floating-point operand: \texttt{gray}

Table 73: ``Set the stroking colour space to DeviceGray (or the DefaultGray colour
space; see 8.6.5.6, "Default Colour Spaces") and set the gray level to use
for stroking operations. gray shall be a number between 0.0 (black) and
1.0 (white).'' \cite{PDF_2020}

\subsubsection{g Operator}
Takes 1 floating-point operand: \texttt{gray}

Table 73: ``Same as G but used for nonstroking operations.'' \cite{PDF_2020}

\begin{figure}[h]
\centering
\begin{subfigure}{0.45\textwidth}
\centering
\includegraphics[width=0.9\linewidth, height=3cm, trim={0 8cm 0 0}, clip]{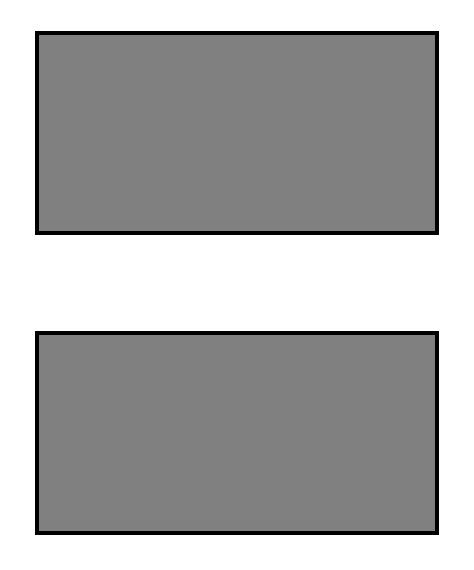}
\caption{\texttt{0.5 g}}
\label{fig:g_subim1}
\end{subfigure}
\begin{subfigure}{0.45\textwidth}
\centering
\includegraphics[width=0.9\linewidth, height=3cm, trim={0 8cm 0 0}, clip]{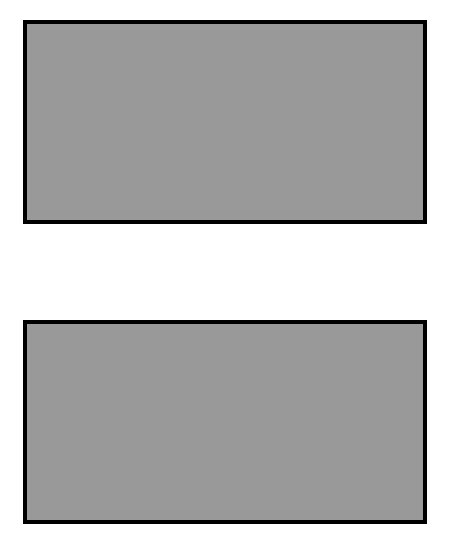}
\caption{\texttt{0.6 g}}
\label{fig:g_subim2}
\end{subfigure}

\begin{subfigure}{0.45\textwidth}
\centering
\includegraphics[width=0.9\linewidth, height=3cm, trim={0 8cm 0 0}, clip]{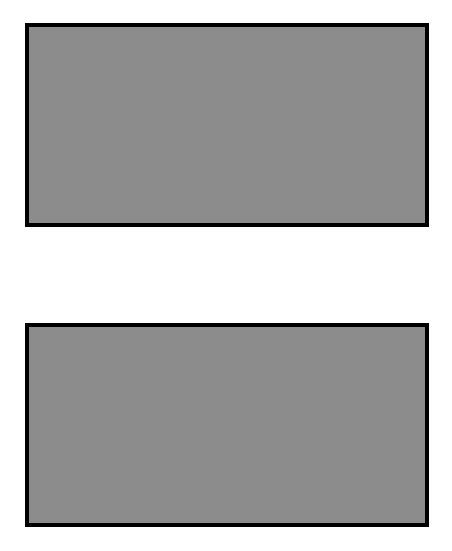}
\caption{\texttt{0.55 g}}
\label{fig:g_subim3}
\end{subfigure}
\begin{subfigure}{0.45\textwidth}
\centering
\includegraphics[width=0.9\linewidth, height=3cm]{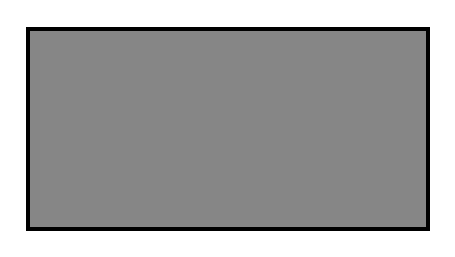}
\caption{\texttt{0.525 g}}
\label{fig:g_subim4}
\end{subfigure}

\begin{subfigure}{0.45\textwidth}
\centering
\includegraphics[width=0.9\linewidth, height=3cm]{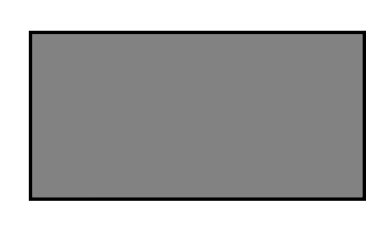}
\caption{\texttt{0.51 g}}
\label{fig:g_subim5}
\end{subfigure}
\begin{subfigure}{0.45\textwidth}
\centering
\includegraphics[width=0.9\linewidth, height=3cm, trim={0 8cm 0 0}, clip]{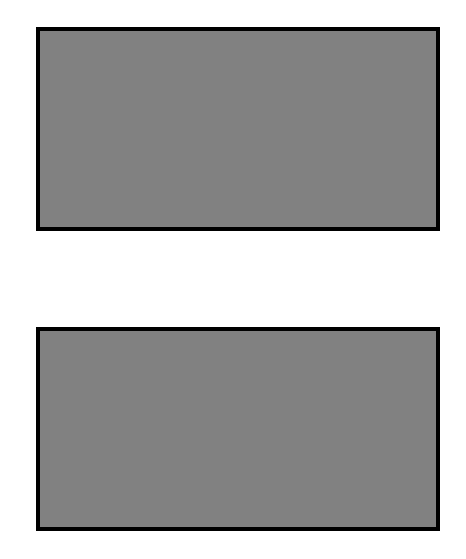}
\caption{\texttt{0.505 g}}
\label{fig:g_subim6}
\end{subfigure}
\caption{g Operator - control, 20\%, 10\%, 5\%, 2\%, 1\% (5\%)}
\label{fig:g_op}
\end{figure}

\subsubsection{K Operator}
Takes 4 floating-point operands: \texttt{c m y k}

Table 73: ``Set the stroking colour space to DeviceCMYK (or the DefaultCMYK
colour space; see 8.6.5.6, "Default Colour Spaces") and set the colour to
use for stroking operations. Each operand shall be a number between 0.0
(zero concentration) and 1.0 (maximum concentration). The behaviour of
this operator is affected by the overprint mode (see 8.6.7, "Overprint
Control").'' \cite{PDF_2020}

\subsubsection{k Operator}
Takes 4 floating-point operands: \texttt{c m y k}

Table 73: ``Same as K but used for nonstroking operations.'' \cite{PDF_2020}

\begin{figure}[h]
\centering
\begin{subfigure}{0.45\textwidth}
\centering
\includegraphics[width=0.9\linewidth, height=3cm]{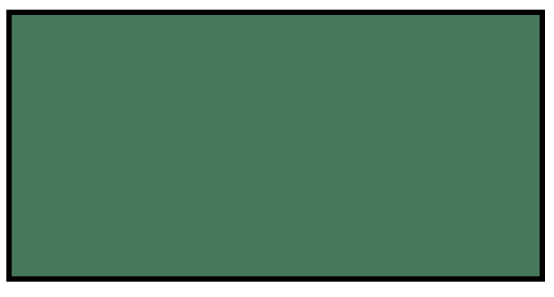}
\caption{\texttt{0.5 0 0.5 0.5 k}}
\label{fig:k_subim1}
\end{subfigure}
\begin{subfigure}{0.45\textwidth}
\centering
\includegraphics[width=0.9\linewidth, height=3cm]{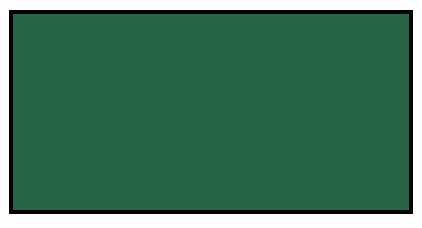}
\caption{\texttt{0.6 0 0.6 0.6 k}}
\label{fig:k_subim2}
\end{subfigure}

\begin{subfigure}{0.45\textwidth}
\centering
\includegraphics[width=0.9\linewidth, height=3cm]{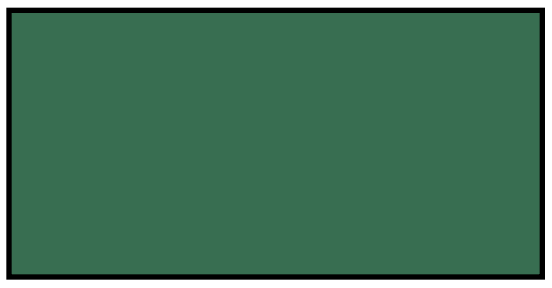}
\caption{\texttt{0.55 0 0.55 0.55 k}}
\label{fig:k_subim3}
\end{subfigure}
\begin{subfigure}{0.45\textwidth}
\centering
\includegraphics[width=0.9\linewidth, height=3cm]{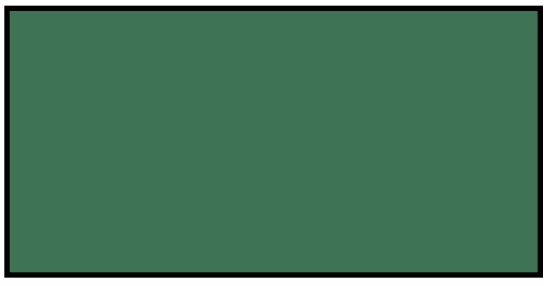}
\caption{\texttt{0.525 0 0.525 0.525 k}}
\label{fig:k_subim4}
\end{subfigure}

\begin{subfigure}{0.45\textwidth}
\centering
\includegraphics[width=0.9\linewidth, height=3cm]{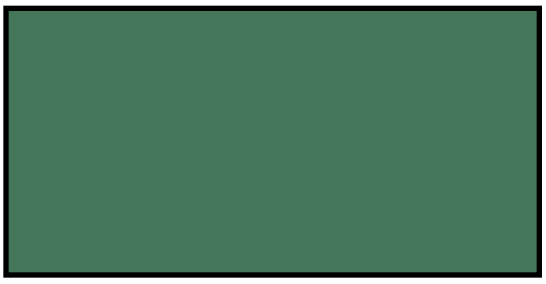}
\caption{\texttt{0.51 0 0.51 0.51 k}}
\label{fig:k_subim5}
\end{subfigure}
\begin{subfigure}{0.45\textwidth}
\centering
\includegraphics[width=0.9\linewidth, height=3cm]{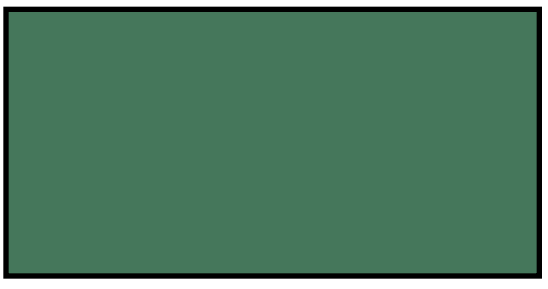}
\caption{\texttt{0.505 0 0.505 0.505 k}}
\label{fig:k_subim6}
\end{subfigure}
\caption{k Operator - control, 20\%, 10\%, 5\%, 2\%, 1\% (5\%)}
\label{fig:k_op}
\end{figure}

\subsubsection{RG Operator}
Takes 3 floating-point operands: \texttt{r g b}

Table 73: ``Set the stroking colour space to DeviceRGB (or the DefaultRGB colour
space; see 8.6.5.6, "Default Colour Spaces") and set the colour to use for
stroking operations. Each operand shall be a number between 0.0
(minimum intensity) and 1.0 (maximum intensity).'' \cite{PDF_2020}

\subsubsection{rg Operator}
Takes 3 floating-point operands: \texttt{r g b}

Table 73: ``Same as RG but used for nonstroking operations.'' \cite{PDF_2020}

\begin{figure}[h]
\centering
\begin{subfigure}{0.45\textwidth}
\centering
\includegraphics[width=0.9\linewidth, height=3cm, trim={0 8cm 0 0}, clip]{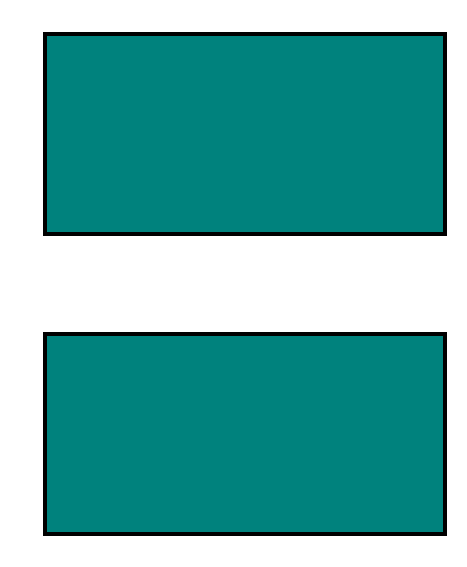}
\caption{\texttt{0 0.5 0.5 rg}}
\label{fig:rg_subim1}
\end{subfigure}
\begin{subfigure}{0.45\textwidth}
\centering
\includegraphics[width=0.9\linewidth, height=3cm, trim={0 8cm 0 0}, clip]{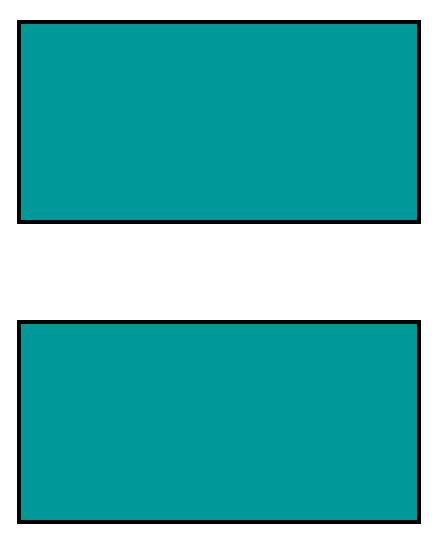}
\caption{\texttt{0 0.6 0.6 rg}}
\label{fig:rg_subim2}
\end{subfigure}

\begin{subfigure}{0.45\textwidth}
\centering
\includegraphics[width=0.9\linewidth, height=3cm, trim={0 8cm 0 0}, clip]{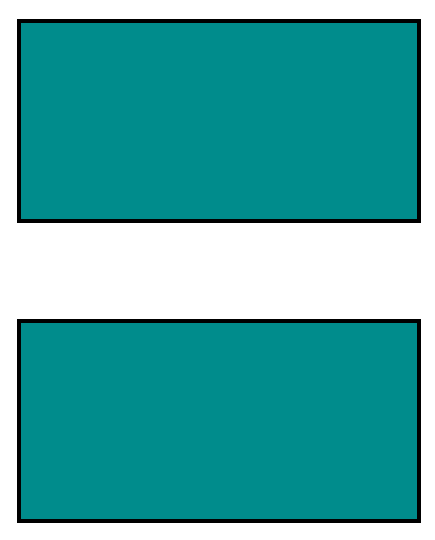}
\caption{\texttt{0 0.55 0.55 rg}}
\label{fig:rg_subim3}
\end{subfigure}
\begin{subfigure}{0.45\textwidth}
\centering
\includegraphics[width=0.9\linewidth, height=3cm]{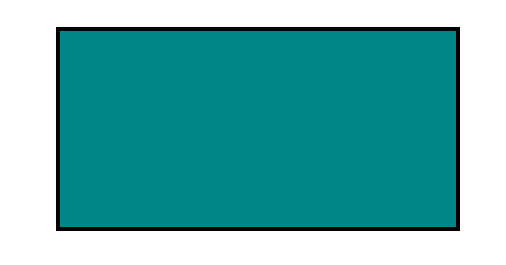}
\caption{\texttt{0 0.525 0.525 rg}}
\label{fig:rg_subim4}
\end{subfigure}

\begin{subfigure}{0.45\textwidth}
\centering
\includegraphics[width=0.9\linewidth, height=3cm]{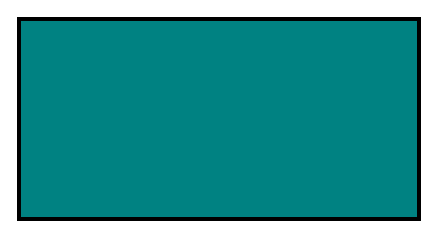}
\caption{\texttt{0 0.51 0.51 rg}}
\label{fig:rg_subim5}
\end{subfigure}
\begin{subfigure}{0.45\textwidth}
\centering
\includegraphics[width=0.9\linewidth, height=3cm, trim={0 8cm 0 0}, clip]{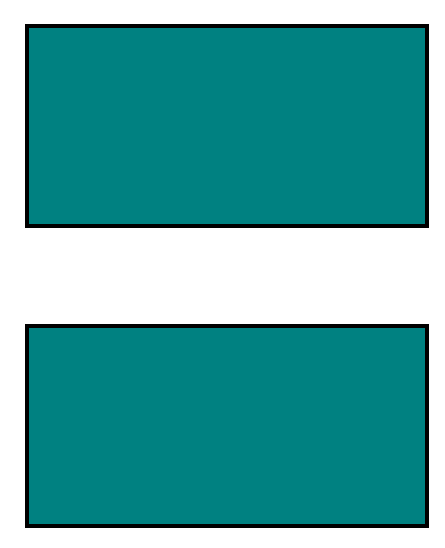}
\caption{\texttt{0 0.505 0.505 rg}}
\label{fig:rg_subim6}
\end{subfigure}
\caption{rg Operator - control, 20\%, 10\%, 5\%, 2\%, 1\% (5\%)}
\label{fig:rg_op}
\end{figure}

\subsubsection{sc Operator}
Takes 1+ floating-point operands: \texttt{$c_1 ... c_n$}

Table 73: ``(PDF 1.1) Same as SC but used for nonstroking operations.''\cite{PDF_2020}\\

Thus when $n = 1$, \texttt{sc} family of operators are equivalent to \texttt{G} and \texttt{g} operators.\\
When $n = 3$, \texttt{sc} family of operators are equivalent to \texttt{RG} and \texttt{rg} operators.\\
When $n = 4$, \texttt{sc} family of operators are equivalent to \texttt{K} and \texttt{k} operators.\\
Thus, since the cutoff for all operators \texttt{G, g, RG, rg, K, k} is 5\%, the cutoff for all operators in the \texttt{sc} family must be 5\%.

\subsubsection{SC Operator}
Takes 1+ floating-point operands: \texttt{$c_1 ... c_n$}

Table 73: ``(PDF 1.1) Set the colour to use for stroking operations in a device, CIE-
based (other than ICCBased), or Indexed colour space. The number of
operands required and their interpretation depends on the current
stroking colour space:\\
For DeviceGray, CalGray, and Indexed colour spaces, one operand
shall be required (n = 1).\\
For DeviceRGB, CalRGB, and Lab colour spaces, three operands shall
be required (n = 3).\\
For DeviceCMYK, four operands shall be required (n = 4).'' \cite{PDF_2020}

\subsubsection{scn Operator}
Takes 1+ floating-point operands: \texttt{$c_1 ... c_n$}

Table 73: ``(PDF 1.2) Same as SCN but used for nonstroking operations.'' \cite{PDF_2020}

\subsubsection{SCN Operator}
Takes 1+ floating-point operands: \texttt{$c_1 ... c_n$}

Table 73: ``(PDF 1.2) Same as SC but also supports Pattern, Separation, DeviceN
and ICCBased colour spaces.'' \cite{PDF_2020}

\subsection{Text Control Operators}
The following operators control how text is placed inside of a PDF document. Many take units in terms of `text units', either scaled by some value or unscaled. In determining perceptibly of changes, it is important to consider both the vertical and horizontal changes in text scaling.

\subsubsection{Tc Operator}
Takes 1 floating-point operand: \texttt{charSpace}

Table 103: ``Set the character spacing, $T_c$ , to charSpace, which shall be a number
expressed in unscaled text space units. Character spacing shall be used
by the Tj, TJ, and ' operators. Initial value: 0.'' \cite{PDF_2020}

\clearpage 

\begin{figure}[h]
\centering
\begin{subfigure}{0.45\textwidth}
\centering
\includegraphics[width=0.9\linewidth, height=1.5cm]{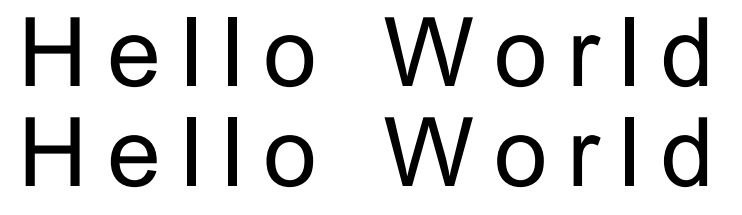}
\caption{\texttt{5 Tc \\ 5 Tc}}
\label{fig:tc_subim1}
\end{subfigure}
\begin{subfigure}{0.45\textwidth}
\centering
\includegraphics[width=0.9\linewidth, height=1.5cm]{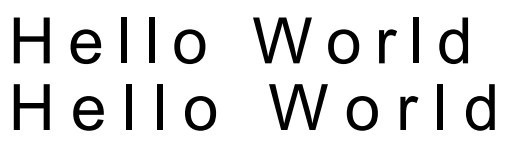}
\caption{\texttt{5 Tc \\ 6 Tc}}
\label{fig:tc_subim2}
\end{subfigure}

\begin{subfigure}{0.45\textwidth}
\centering
\includegraphics[width=0.9\linewidth, height=1.5cm]{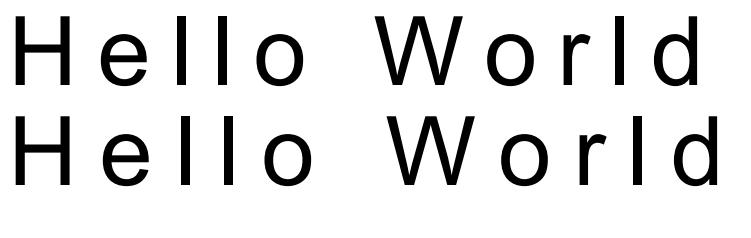}
\caption{\texttt{5 Tc \\ 5.5 Tc}}
\label{fig:tc_subim3}
\end{subfigure}
\begin{subfigure}{0.45\textwidth}
\centering
\includegraphics[width=0.9\linewidth, height=1.5cm]{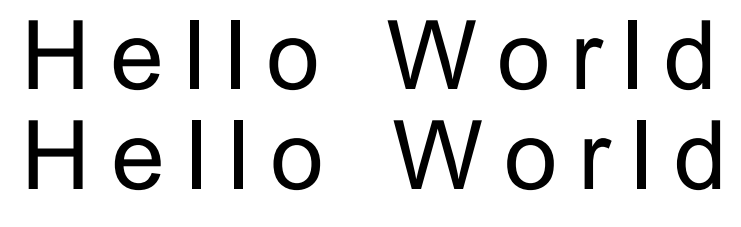}
\caption{\texttt{5 Tc \\ 5.25 Tc}}
\label{fig:tc_subim4}
\end{subfigure}

\begin{subfigure}{0.45\textwidth}
\centering
\includegraphics[width=0.9\linewidth, height=1.5cm]{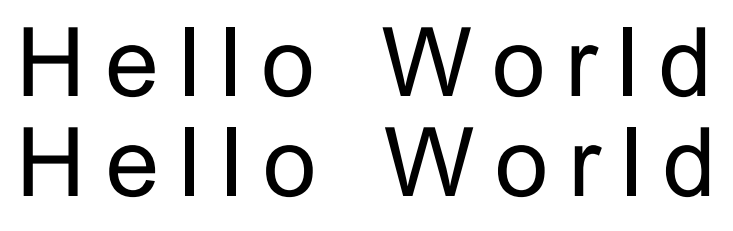}
\caption{\texttt{5 Tc \\ 5.1 Tc}}
\label{fig:tc_subim5}
\end{subfigure}
\begin{subfigure}{0.45\textwidth}
\centering
\includegraphics[width=0.9\linewidth, height=1.5cm]{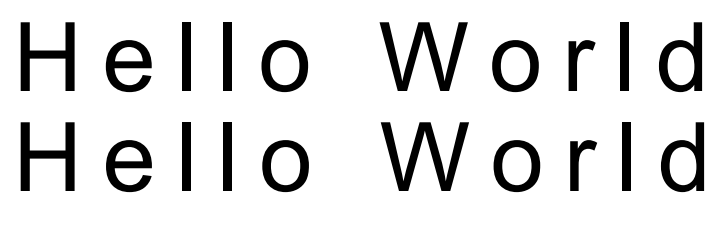}
\caption{\texttt{5 Tc \\ 5.05 Tc}}
\label{fig:tc_subim6}
\end{subfigure}
\caption{Tc Operator - control, 20\%, 10\%, 5\%, 2\%, 1\% (1\%, less stable)}
\label{fig:tc_op}
\end{figure}

\subsubsection{Td Operator}
Takes 1 floating-point operand: \texttt{$t_x \ t_y$}

Table 106: ``Move to the start of the next line, offset from the start of the current line by
($t_x$ , $t_y$ ). $t_x$ and $t_y$ shall denote numbers expressed in unscaled text space
units.'' \cite{PDF_2020}

\clearpage 

\begin{figure}[h]
\centering
\begin{subfigure}{0.45\textwidth}
\centering
\includegraphics[width=0.9\linewidth, height=1.5cm]{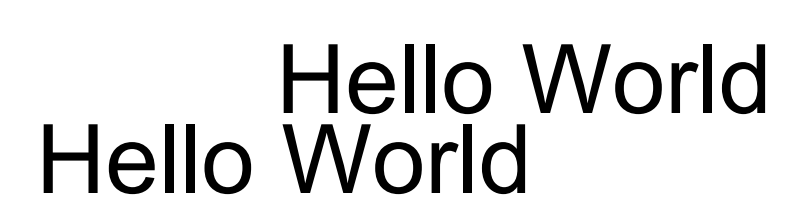}
\caption{\texttt{60 20 Td}}
\label{fig:td_subim1}
\end{subfigure}
\begin{subfigure}{0.45\textwidth}
\centering
\includegraphics[width=0.9\linewidth, height=1.5cm]{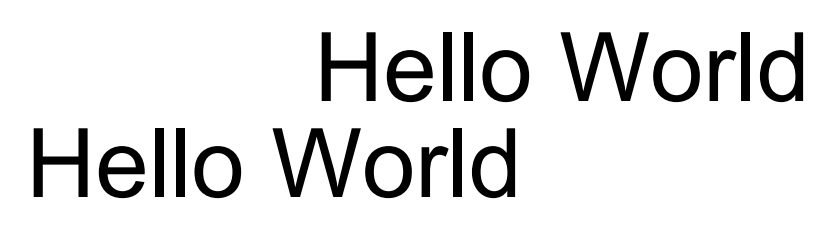}
\caption{\texttt{72 24 Td}}
\label{fig:td_subim2}
\end{subfigure}

\begin{subfigure}{0.45\textwidth}
\centering
\includegraphics[width=0.9\linewidth, height=1.5cm]{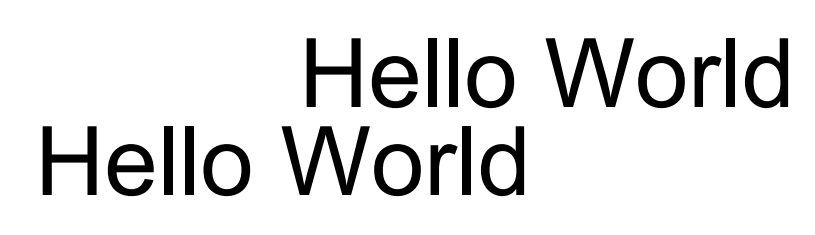}
\caption{\texttt{66 22 Td}}
\label{fig:td_subim3}
\end{subfigure}
\begin{subfigure}{0.45\textwidth}
\centering
\includegraphics[width=0.9\linewidth, height=1.5cm]{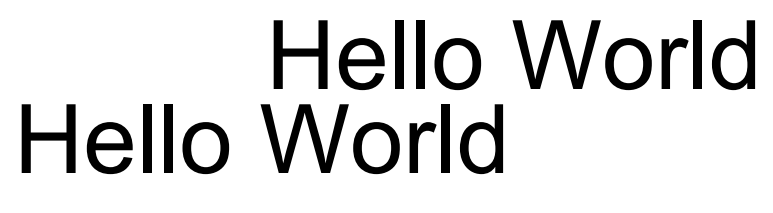}
\caption{\texttt{63 21 Td}}
\label{fig:td_subim4}
\end{subfigure}

\begin{subfigure}{0.45\textwidth}
\centering
\includegraphics[width=0.9\linewidth, height=1.5cm]{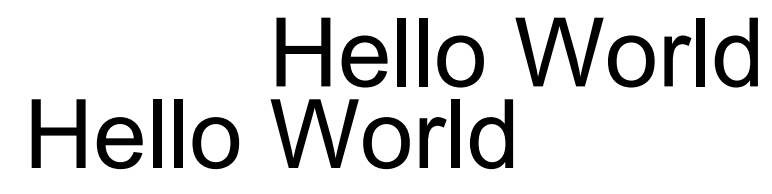}
\caption{\texttt{61.2 20.4 Td}}
\label{fig:td_subim5}
\end{subfigure}
\begin{subfigure}{0.45\textwidth}
\centering
\includegraphics[width=0.9\linewidth, height=1.5cm]{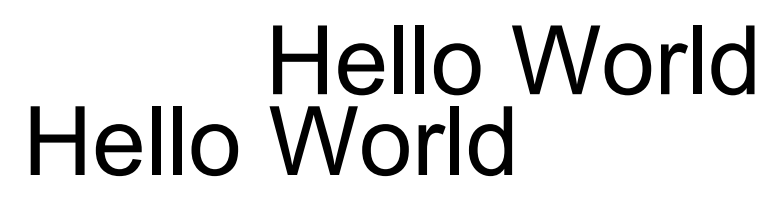}
\caption{\texttt{60.6 20.2 Td}}
\label{fig:td_subim6}
\end{subfigure}
\caption{Td Operator - control, 20\%, 10\%, 5\%, 2\%, 1\% (2\%)}
\label{fig:td_op}
\end{figure}

\subsubsection{TD Operator}
Takes 1 floating-point operand: \texttt{$t_x \ t_y$}

Table 106: ``Move to the start of the next line, offset from the start of the current line by
($t_x$ , $t_y$ ). As a side effect, this operator shall set the leading parameter in
the text state.'' \cite{PDF_2020}

According to the PDF standard, \texttt{x y TD} is equivalent to: \\
\texttt{-y TL \\ x y Td} \cite{PDF_2020} \\

\clearpage 

\begin{figure}[h]
\centering
\begin{subfigure}{0.45\textwidth}
\centering
\includegraphics[width=0.9\linewidth, height=1.75cm]{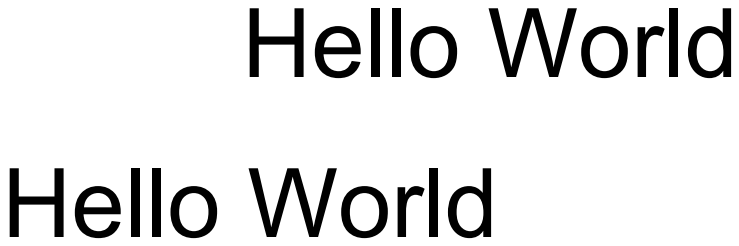}
\caption{\texttt{60 20 TD}}
\label{fig:TD_subim1}
\end{subfigure}
\begin{subfigure}{0.45\textwidth}
\centering
\includegraphics[width=0.9\linewidth, height=1.75cm]{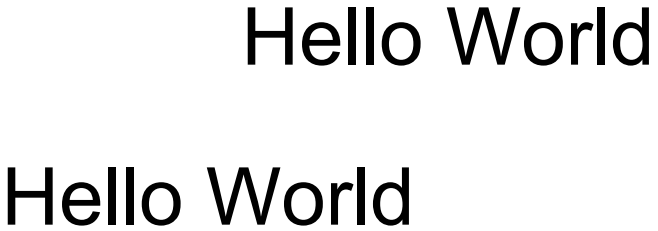}
\caption{\texttt{72 24 TD}}
\label{fig:TD_subim2}
\end{subfigure}

\begin{subfigure}{0.45\textwidth}
\centering
\includegraphics[width=0.9\linewidth, height=1.75cm]{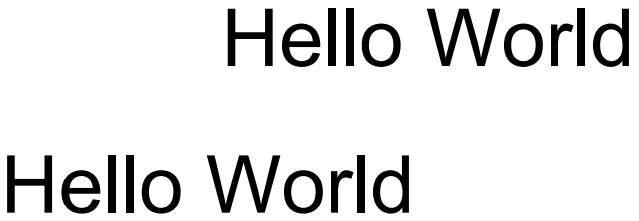}
\caption{\texttt{66 22 TD}}
\label{fig:TD_subim3}
\end{subfigure}
\begin{subfigure}{0.45\textwidth}
\centering
\includegraphics[width=0.9\linewidth, height=1.75cm]{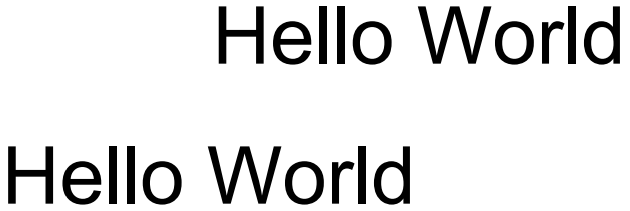}
\caption{\texttt{63 21 TD}}
\label{fig:TD_subim4}
\end{subfigure}

\begin{subfigure}{0.45\textwidth}
\centering
\includegraphics[width=0.9\linewidth, height=1.75cm]{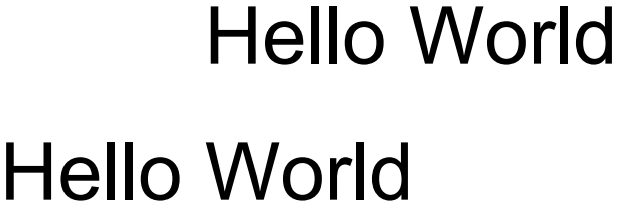}
\caption{\texttt{61.2 20.4 TD}}
\label{fig:TD_subim5}
\end{subfigure}
\begin{subfigure}{0.45\textwidth}
\centering
\includegraphics[width=0.9\linewidth, height=1.75cm]{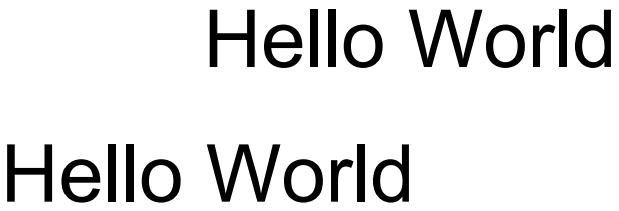}
\caption{\texttt{60.6 20.2 TD}}
\label{fig:TD_subim6}
\end{subfigure}
\caption{TD Operator - control, 20\%, 10\%, 5\%, 2\%, 1\% (2\%)}
\label{fig:TD_op}
\end{figure}

\subsubsection{Tf Operator}
Takes 1 floating-point operand: \texttt{size}

Table 103: ``Set the text font, $T_f$ , to font and the text font size, $T_{fs}$ , to size. font shall be
the name of a font resource in the Font subdictionary of the current
resource dictionary; size shall be a number representing a scale factor.
There is no initial value for either font or size; they shall be specified
explicitly by using Tf before any text is shown.'' \cite{PDF_2020}

\clearpage

\begin{figure}[ht]
\centering
\begin{subfigure}{0.45\textwidth}
\centering
\includegraphics[width=0.9\linewidth, height=2cm]{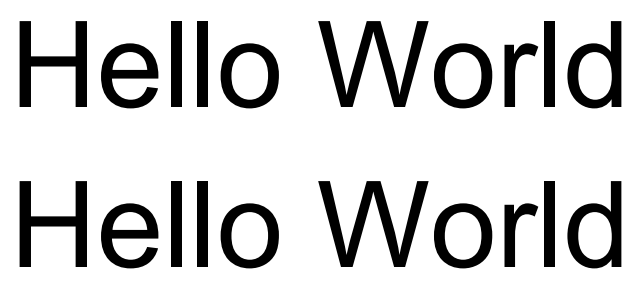}
\caption{\texttt{30 Tf \\ 30 Tf}}
\label{fig:tf_subim1}
\end{subfigure}
\begin{subfigure}{0.45\textwidth}
\centering
\includegraphics[width=0.9\linewidth, height=2cm]{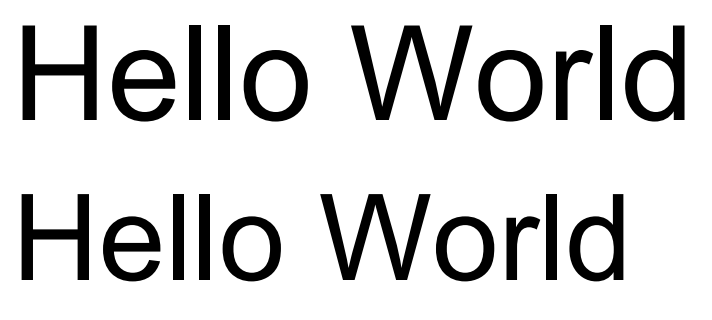}
\caption{\texttt{33 Tf \\ 30 Tf}}
\label{fig:tf_subim2}
\end{subfigure}

\begin{subfigure}{0.45\textwidth}
\centering
\includegraphics[width=0.9\linewidth, height=2cm]{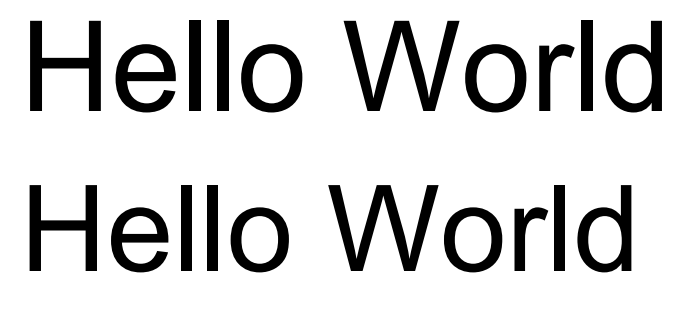}
\caption{\texttt{31.5 Tf \\ 30 Tf}}
\label{fig:tf_subim3}
\end{subfigure}
\begin{subfigure}{0.45\textwidth}
\centering
\includegraphics[width=0.9\linewidth, height=2cm]{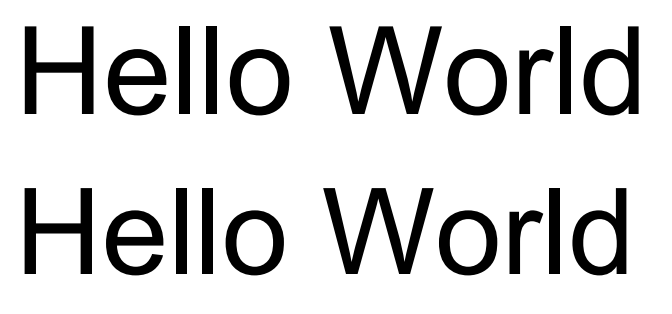}
\caption{\texttt{30.6 Tf \\ 30 Tf}}
\label{fig:tf_subim4}
\end{subfigure}

\begin{subfigure}{0.45\textwidth}
\centering
\includegraphics[width=0.9\linewidth, height=2cm]{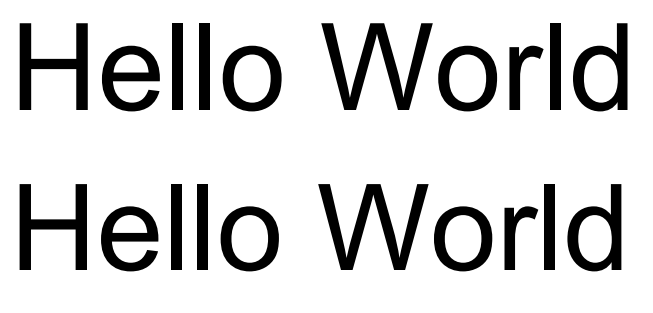}
\caption{\texttt{30.3 Tf \\ 30 Tf}}
\label{fig:tf_subim5}
\end{subfigure}
\begin{subfigure}{0.45\textwidth}
\centering
\includegraphics[width=0.9\linewidth, height=2cm]{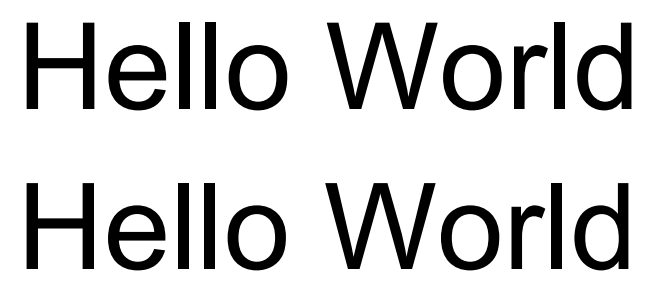}
\caption{\texttt{30.15 Tf \\ 30 Tf}}
\label{fig:tf_subim6}
\end{subfigure}

\begin{subfigure}{0.45\textwidth}
\centering
\includegraphics[width=0.9\linewidth, height=2cm]{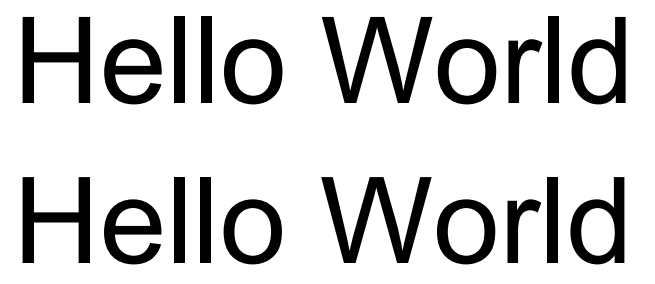}
\caption{\texttt{30.06 Tf \\ 30 Tf}}
\label{fig:tf_subim7}
\end{subfigure}
\begin{subfigure}{0.45\textwidth}
\centering
\includegraphics[width=0.9\linewidth, height=2cm]{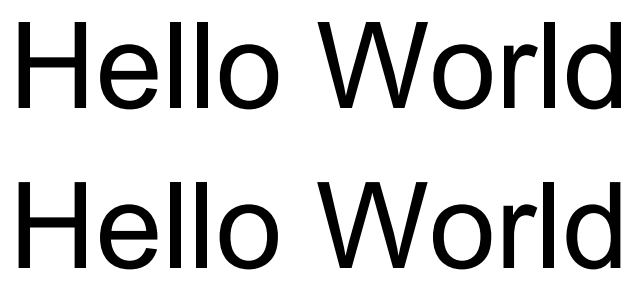}
\caption{\texttt{30.03 Tf \\ 30 Tf}}
\label{fig:tf_subim8}
\end{subfigure}

\caption{Tf Operator - control, 10\%, 5\%, 2\%, 1\%, 0.5\%, 0.2\%, 0.1\% (0.5\%)}
\label{fig:tf_op}
\end{figure}

\subsubsection{TL Operator}
Takes 1 floating-point operand: \texttt{leading}

Table 103: ``Set the text leading, $T_l$ , to leading, which shall be a number expressed in
unscaled text space units. Text leading shall be used only by the T*, ', and
" operators. Initial value: 0.'' \cite{PDF_2020}

\clearpage

\begin{figure}[ht]
\centering
\begin{subfigure}{0.45\textwidth}
\centering
\includegraphics[width=0.9\linewidth, height=2.25cm]{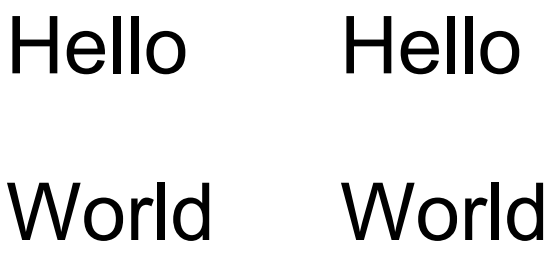}
\caption{\texttt{50 TL \\ 50 TL}}
\label{fig:tl_subim1}
\end{subfigure}
\begin{subfigure}{0.45\textwidth}
\centering
\includegraphics[width=0.9\linewidth, height=2.25cm]{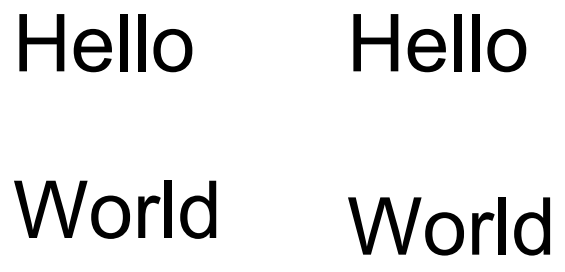}
\caption{\texttt{50 TL \\ 55 TL}}
\label{fig:tl_subim2}
\end{subfigure}

\begin{subfigure}{0.45\textwidth}
\centering
\includegraphics[width=0.9\linewidth, height=2.25cm]{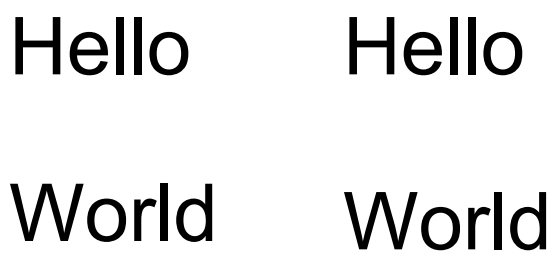}
\caption{\texttt{50 TL \\ 52.5 TL}}
\label{fig:tl_subim3}
\end{subfigure}
\begin{subfigure}{0.45\textwidth}
\centering
\includegraphics[width=0.9\linewidth, height=2.25cm]{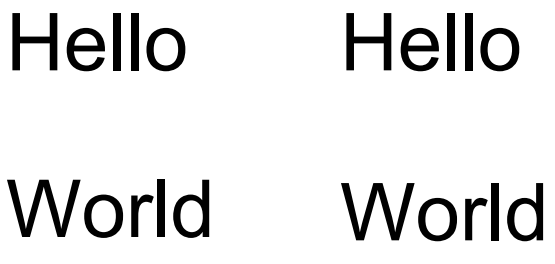}
\caption{\texttt{50 TL \\ 51 TL}}
\label{fig:tl_subim4}
\end{subfigure}

\begin{subfigure}{0.45\textwidth}
\centering
\includegraphics[width=0.9\linewidth, height=2.25cm]{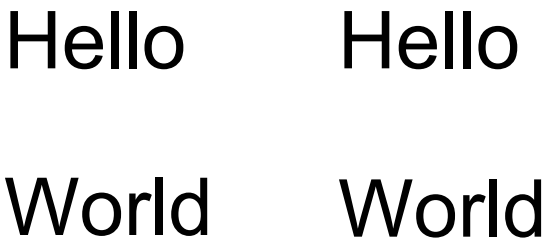}
\caption{\texttt{50 TL \\ 50.5 TL}}
\label{fig:tl_subim5}
\end{subfigure}
\begin{subfigure}{0.45\textwidth}
\centering
\includegraphics[width=0.9\linewidth, height=2.25cm]{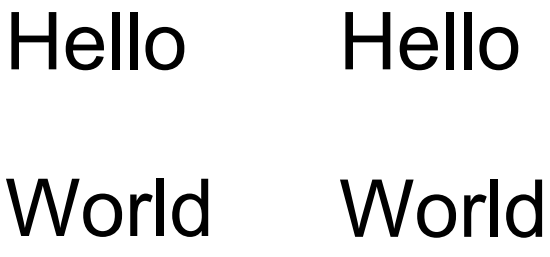}
\caption{\texttt{50 TL \\ 50.25 TL}}
\label{fig:tl_subim6}
\end{subfigure}

\caption{Tl Operator - control, 10\%, 5\%, 2\%, 1\%, 0.5\% (2\%)}
\label{fig:tl_op}
\end{figure}

\subsubsection{Tm Operator}
Takes 1 floating-point operand: \texttt{a b c d e f}

Table 106: ``Set the text matrix, $T_m$ , and the text line matrix, $T_{lm}$ :\\
The operands shall all be numbers, and the initial value for $T_m$ and $T_{lm}$
shall be the identity matrix, [ 1 0 0 1 0 0 ]. Although the operands
specify a matrix, they shall be passed to Tm as six separate numbers, not
as an array.\\
The matrix specified by the operands shall not be concatenated onto the
current text matrix, but shall replace it.'' \cite{PDF_2020}

The \texttt{Tm} operator is similar to the \texttt{cm} graphics operator. Operands \texttt{e} and \texttt{f} move the location where the next piece of text will be drawn. Therefore, those two operands have the same effect as the \texttt{Td} operator, and should have the same percentage cutoff of 2\%. Looking at operands \texttt{a, b, c, d} shows that their percentage cutoff should be 

\clearpage
\begin{figure}[h]
\centering
\begin{subfigure}{0.45\textwidth}
\centering
\includegraphics[width=0.9\linewidth, height=4cm]{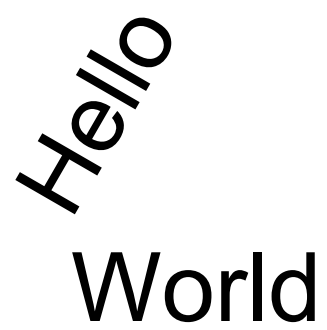}
\caption{\texttt{0.5 0.866 -0.866 0.5 0 0 Tm}}
\label{fig:tm_subim1}
\end{subfigure}
\begin{subfigure}{0.45\textwidth}
\centering
\includegraphics[width=0.9\linewidth, height=4cm]{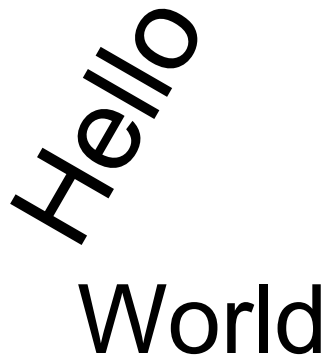}
\caption{\texttt{0.6 1.04 -1.04 0.6 0 0 Tm}}
\label{fig:tm_subim2}
\end{subfigure}

\begin{subfigure}{0.45\textwidth}
\centering
\includegraphics[width=0.9\linewidth, height=4cm]{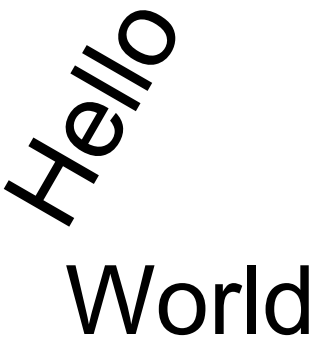}
\caption{\texttt{0.55 0.953 -0.953 0.55 0 0 Tm}}
\label{fig:tm_subim3}
\end{subfigure}
\begin{subfigure}{0.45\textwidth}
\centering
\includegraphics[width=0.9\linewidth, height=4cm]{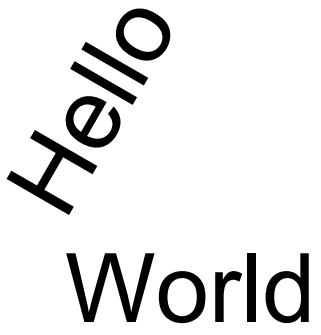}
\caption{\texttt{.525 .909 -.909 .525 0 0 Tm}}
\label{fig:tm_subim4}
\end{subfigure}

\begin{subfigure}{0.45\textwidth}
\centering
\includegraphics[width=0.9\linewidth, height=4cm]{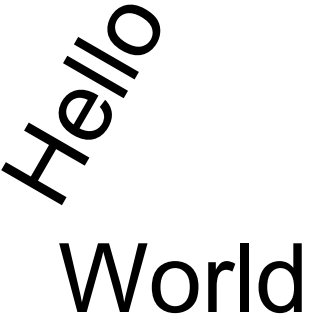}
\caption{\texttt{0.51 0.883 -0.883 0.51 0 0 Tm}}
\label{fig:tm_subim5}
\end{subfigure}
\begin{subfigure}{0.45\textwidth}
\centering
\includegraphics[width=0.9\linewidth, height=4cm]{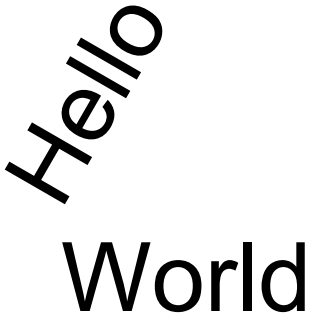}
\caption{\texttt{.505 .875 -.875 .505 0 0 Tm}}
\label{fig:tm_subim6}
\end{subfigure}
\caption{Tm Operator - control, 20\%, 10\%, 5\%, 2\%, 1\% (5-10\%)}
\label{fig:tm_op}
\end{figure}

\clearpage 

\subsubsection{Ts Operator}
Takes 1 floating-point operand: \texttt{rise}

Table 103: ``Set the text rise, $T_{rise}$ , to rise, which shall be a number expressed in
unscaled text space units. Initial value: 0.'' \cite{PDF_2020}

\begin{figure}[h]
\centering
\begin{subfigure}{0.45\textwidth}
\centering
\includegraphics[width=0.9\linewidth, height=1.5cm]{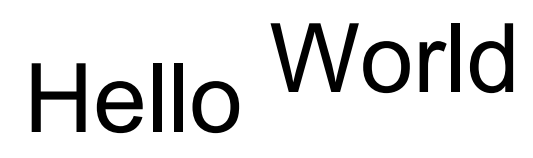}
\caption{\texttt{10 Ts}}
\label{fig:ts_subim1}
\end{subfigure}
\begin{subfigure}{0.45\textwidth}
\centering
\includegraphics[width=0.9\linewidth, height=1.5cm]{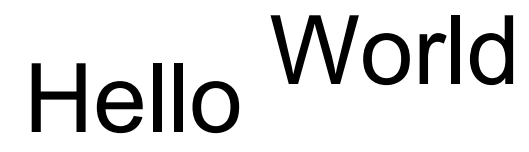}
\caption{\texttt{12 Ts}}
\label{fig:ts_subim2}
\end{subfigure}

\begin{subfigure}{0.45\textwidth}
\centering
\includegraphics[width=0.9\linewidth, height=1.5cm]{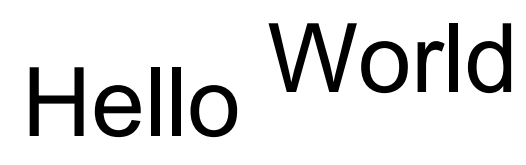}
\caption{\texttt{11 Ts}}
\label{fig:ts_subim3}
\end{subfigure}
\begin{subfigure}{0.45\textwidth}
\centering
\includegraphics[width=0.9\linewidth, height=1.25cm]{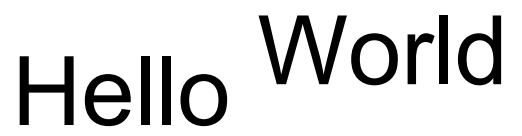}
\caption{\texttt{10.5 Ts}}
\label{fig:ts_subim4}
\end{subfigure}

\begin{subfigure}{0.45\textwidth}
\centering
\includegraphics[width=0.9\linewidth, height=1.5cm]{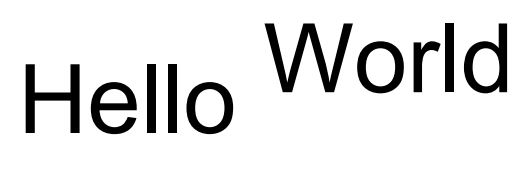}
\caption{\texttt{10.2 Ts}}
\label{fig:ts_subim5}
\end{subfigure}
\begin{subfigure}{0.45\textwidth}
\centering
\includegraphics[width=0.9\linewidth, height=1.5cm]{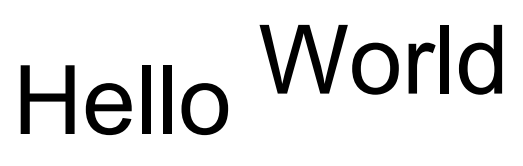}
\caption{\texttt{10.1 Ts}}
\label{fig:ts_subim6}
\end{subfigure}
\caption{Ts Operator - control, 20\%, 10\%, 5\%, 2\%, 1\% (5-10\%)}
\label{fig:ts_op}
\end{figure}

\subsubsection{Tw Operator}
Takes 1 floating-point operand: \texttt{wordSpace}

Table 103: ``Set the word spacing, $T_w$, to wordSpace, which shall be a number
expressed in unscaled text space units. Word spacing shall be used by
the Tj, TJ, and ' operators. Initial value: 0.'' \cite{PDF_2020}

\clearpage

\begin{figure}[h]
\centering
\begin{subfigure}{0.45\textwidth}
\centering
\includegraphics[width=0.9\linewidth, height=1.25cm]{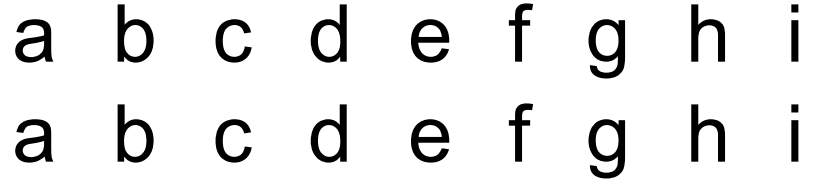}
\caption{\texttt{10 Tw \\ 10 Tw}}
\label{fig:tw_subim1}
\end{subfigure}
\begin{subfigure}{0.45\textwidth}
\centering
\includegraphics[width=0.9\linewidth, height=1.25cm]{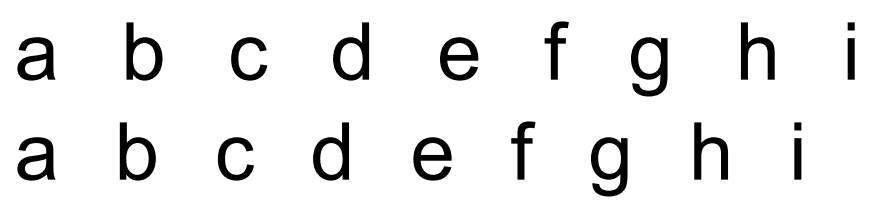}
\caption{\texttt{12 Tw \\ 10 Tw}}
\label{fig:tw_subim2}
\end{subfigure}

\begin{subfigure}{0.45\textwidth}
\centering
\includegraphics[width=0.9\linewidth, height=1.25cm]{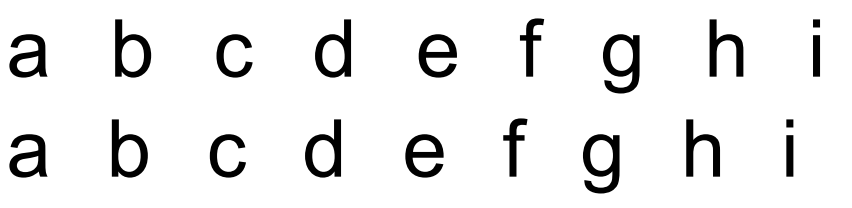}
\caption{\texttt{11 Tw \\ 10 Tw}}
\label{fig:tw_subim3}
\end{subfigure}
\begin{subfigure}{0.45\textwidth}
\centering
\includegraphics[width=0.9\linewidth, height=1.25cm]{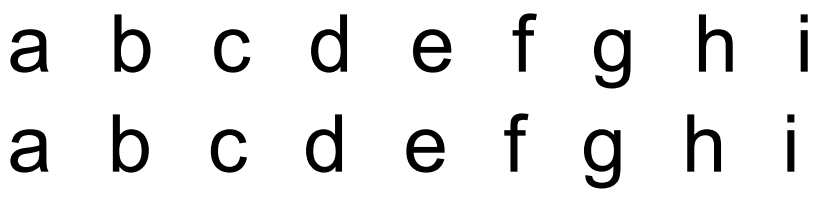}
\caption{\texttt{10.5 Tw \\ 10 Tw}}
\label{fig:tw_subim4}
\end{subfigure}

\begin{subfigure}{0.45\textwidth}
\centering
\includegraphics[width=0.9\linewidth, height=1.25cm]{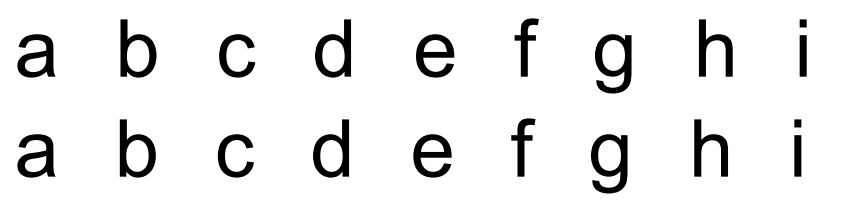}
\caption{\texttt{10.2 Tw \\ 10 Tw}}
\label{fig:tw_subim5}
\end{subfigure}
\begin{subfigure}{0.45\textwidth}
\centering
\includegraphics[width=0.9\linewidth, height=1.25cm]{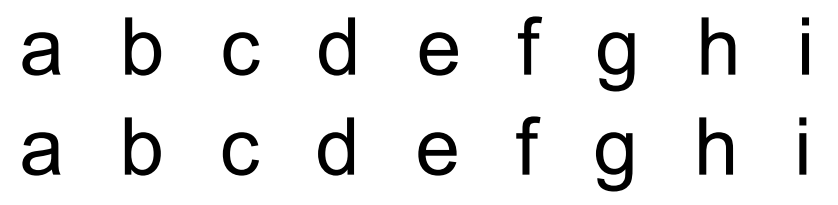}
\caption{\texttt{10.1 Tw \\ 10 Tw}}
\label{fig:tw_subim6}
\end{subfigure}
\caption{Tw Operator - control, 20\%, 10\%, 5\%, 2\%, 1\% (1\%, less stable?)}
\label{fig:tw_op}
\end{figure}

\subsubsection{Tz Operator}
Takes 1 floating-point operand: \texttt{scale}

Table 103: ``Set the horizontal scaling, $T_h$ , to (scale ÷ 100). scale shall be a number
specifying the percentage of the normal width. Initial value: 100 (normal
width).'' \cite{PDF_2020}

\clearpage

\begin{figure}[h]
\centering
\begin{subfigure}{0.45\textwidth}
\centering
\includegraphics[width=0.9\linewidth, height=2.25cm]{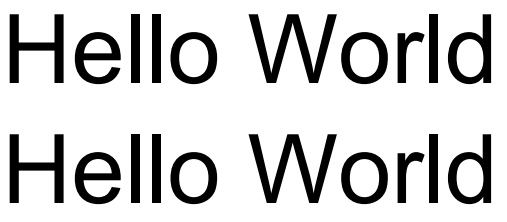}
\caption{\texttt{100 Tz \\ 100 Tz}}
\label{fig:tz_subim1}
\end{subfigure}
\begin{subfigure}{0.45\textwidth}
\centering
\includegraphics[width=0.9\linewidth, height=2.25cm]{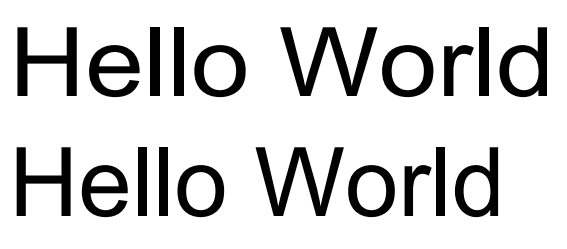}
\caption{\texttt{110 Tz \\ 100 Tz}}
\label{fig:tz_subim2}
\end{subfigure}

\begin{subfigure}{0.45\textwidth}
\centering
\includegraphics[width=0.9\linewidth, height=2.25cm]{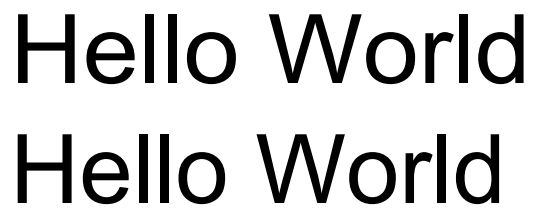}
\caption{\texttt{105 Tz \\ 100 Tz}}
\label{fig:tz_subim3}
\end{subfigure}
\begin{subfigure}{0.45\textwidth}
\centering
\includegraphics[width=0.9\linewidth, height=2.25cm]{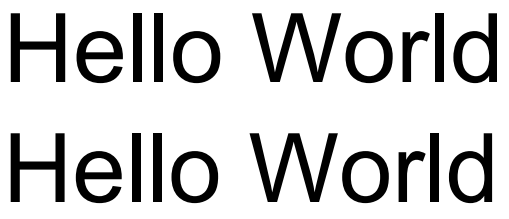}
\caption{\texttt{101 Tz \\ 100 Tz}}
\label{fig:tz_subim4}
\end{subfigure}

\begin{subfigure}{0.45\textwidth}
\centering
\includegraphics[width=0.9\linewidth, height=2.25cm]{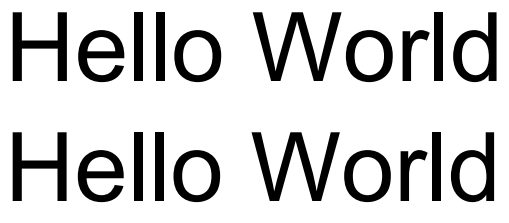}
\caption{\texttt{100.5 Tz \\ 100 Tz}}
\label{fig:tz_subim5}
\end{subfigure}
\begin{subfigure}{0.45\textwidth}
\centering
\includegraphics[width=0.9\linewidth, height=2.25cm]{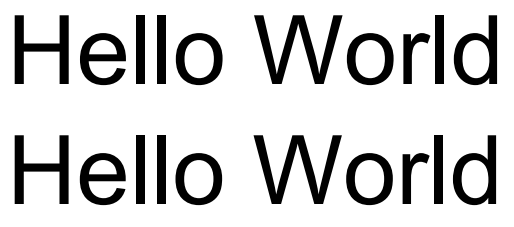}
\caption{\texttt{100.2 Tz \\ 100 Tz}}
\label{fig:tz_subim6}
\end{subfigure}
\caption{Tz Operator - control,  10\%, 5\%, 1\%, 0.5\%, 0.2\% (0.5\%)}
\label{fig:tz_op}
\end{figure}

\subsubsection{TJ Operator}
Takes an array of string and numerical operands. The numerical operands have been proven in previous work (Papers 9.1.1, 9.1.2) to be viable for steganography. Since the units are scaled to thousandths of text space units, a very significant percentage change is necessary to create a noticeable visual change. Extensive study has been done on using this operator for steganography. While our method will not likely yield improved embedding rates for the \texttt{TJ} operator in particular, . 

Table 107: ``Show one or more text strings, allowing individual glyph positioning. Each
element of array shall be either a string or a number. If the element is a
string, this operator shall show the string. If it is a number, the operator
shall adjust the text position by that amount; that is, it shall translate the
text matrix, Tm. The number shall be expressed in thousandths of a unit
of text space (see 9.4.4, "Text Space Details"). This amount shall be
subtracted from the current horizontal or vertical coordinate, depending
on the writing mode. In the default coordinate system, a positive
adjustment has the effect of moving the next glyph painted either to the
left or down by the given amount''. \cite{PDF_2020}

\begin{figure}[h]
\centering
\begin{subfigure}{0.45\textwidth}
\centering
\includegraphics[width=0.4\linewidth, height=2.5cm]{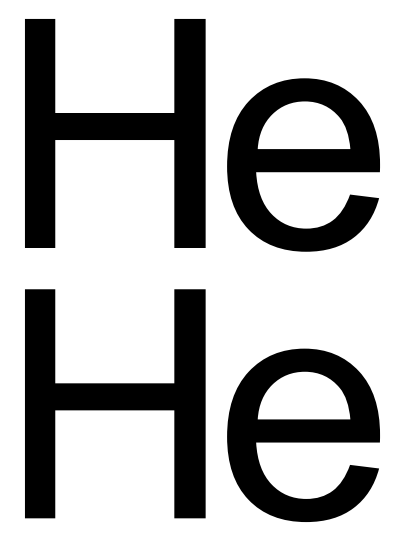}
\caption{\texttt{{[(H)50(e)]} TJ \\ $ \ \ $ {[(H)50(e)]} TJ}}
\label{fig:TJ_subim1}
\end{subfigure}
\begin{subfigure}{0.45\textwidth}
\centering 
\includegraphics[width=0.4\linewidth, height=2.5cm]{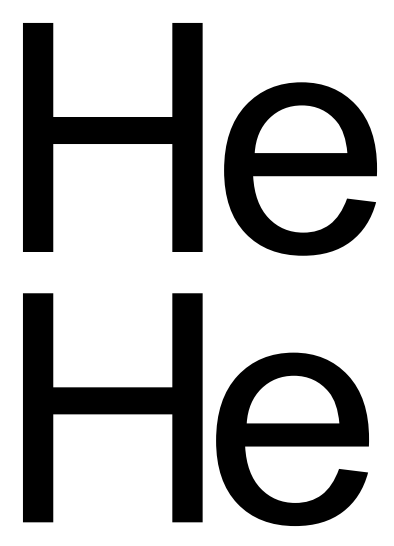}
\caption{\texttt{{[(H)50(e)]} TJ \\ $ \ \ $ {[(H)75(e)]} TJ}}
\label{fig:TJ_subim2}
\end{subfigure}

\begin{subfigure}{0.45\textwidth}
\centering
\includegraphics[width=0.4\linewidth, height=2.5cm]{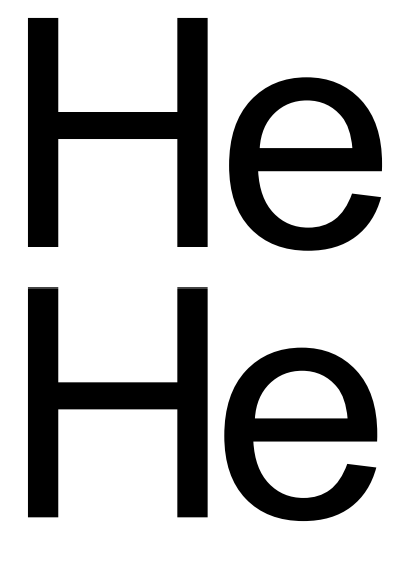}
\caption{\texttt{{[(H)50(e)]} TJ \\ $ \ \ $ {[(H)65(e)]} TJ}}
\label{fig:TJ_subim3}
\end{subfigure}
\begin{subfigure}{0.45\textwidth}
\centering
\includegraphics[width=0.4\linewidth, height=2.5cm]{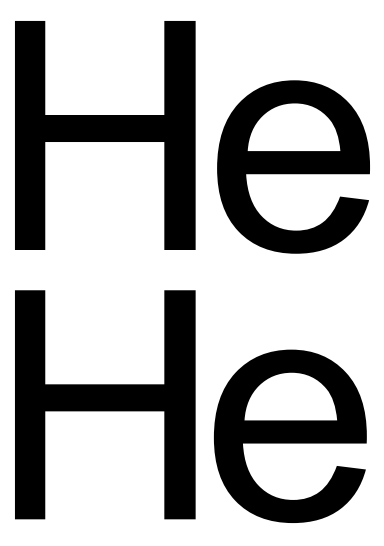}
\caption{\texttt{{[(H)50(e)]} TJ \\ $ \ \ $ {[(H)60(e)]} TJ}}
\label{fig:TJ_subim4}
\end{subfigure}

\begin{subfigure}{0.45\textwidth}
\centering
\includegraphics[width=0.4\linewidth, height=2.5cm]{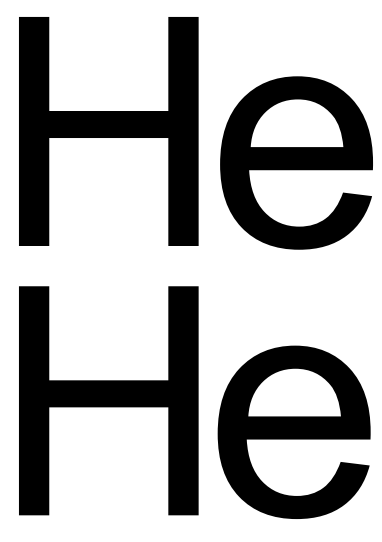}
\caption{\texttt{{[(H)50(e)]} TJ \\  {[(H)57.5(e)]} TJ}}
\label{fig:TJ_subim5}
\end{subfigure}
\begin{subfigure}{0.45\textwidth}
\centering
\includegraphics[width=0.4\linewidth, height=2.5cm]{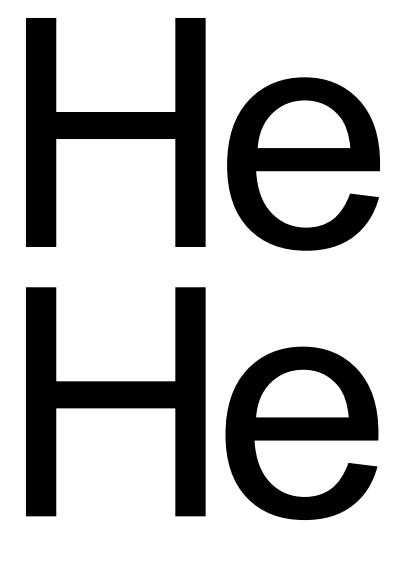}
\caption{\texttt{{[(H)50(e)]} TJ \\ $ \ \ $ {[(H)55(e)]} TJ}}
\label{fig:TJ_subim6}
\end{subfigure}
\caption{TJ Operator - control,  50\%, 30\%, 20\%, 15\%, 10\% (20\%)}
\label{fig:TJ_op}
\end{figure}

\subsection{Misc. Operators}
Note: not deprecated, just very infrequently used in practice. But it is still in the new standard.

\subsubsection{d0 Operator}
Takes 2 floating-point operands: \texttt{$w_x \ w_y$}

Table 111: ``$w_x$ denotes the horizontal displacement in the glyph coordinate
system; it shall be consistent with the corresponding width in the
font’s Widths array. $w_y$ shall be 0 (see 9.2.4, "Glyph Positioning
and Metrics").'' \cite{PDF_2020}

Therefore, it effectively only has 1 operand, since $w_y$ must be fixed to 0

\clearpage

\begin{figure}[h]
\centering
\begin{subfigure}{0.45\textwidth}
\centering
\includegraphics[width=0.9\linewidth, height=4.25cm]{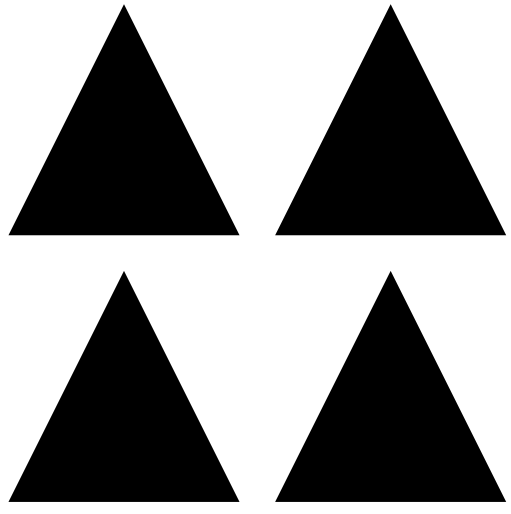}
\caption{\texttt{1000 0 d0 \\ 1000 0 d0}}
\label{fig:d0_subim1}
\end{subfigure}
\begin{subfigure}{0.45\textwidth}
\centering
\includegraphics[width=0.9\linewidth, height=4.25cm]{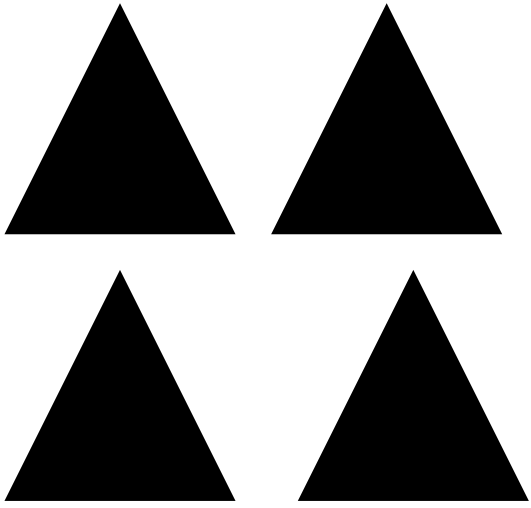}
\caption{\texttt{1000 0 d0 \\ 1100 0 d0}}
\label{fig:d0_subim2}
\end{subfigure}

\begin{subfigure}{0.45\textwidth}
\centering
\includegraphics[width=0.9\linewidth, height=4.25cm]{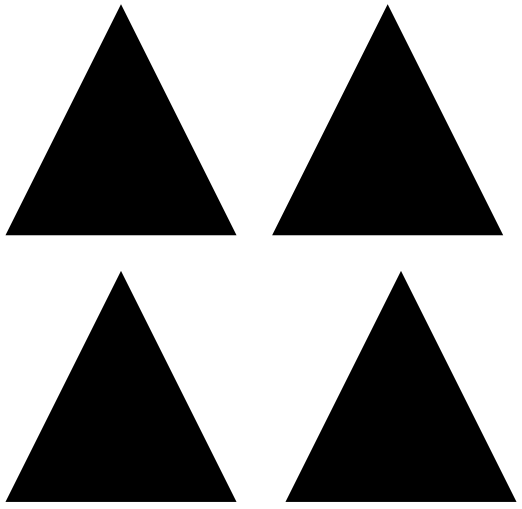}
\caption{\texttt{1000 0 d0 \\ 1050 0 d0}}
\label{fig:d0_subim3}
\end{subfigure}
\begin{subfigure}{0.45\textwidth}
\centering
\includegraphics[width=0.9\linewidth, height=4.25cm]{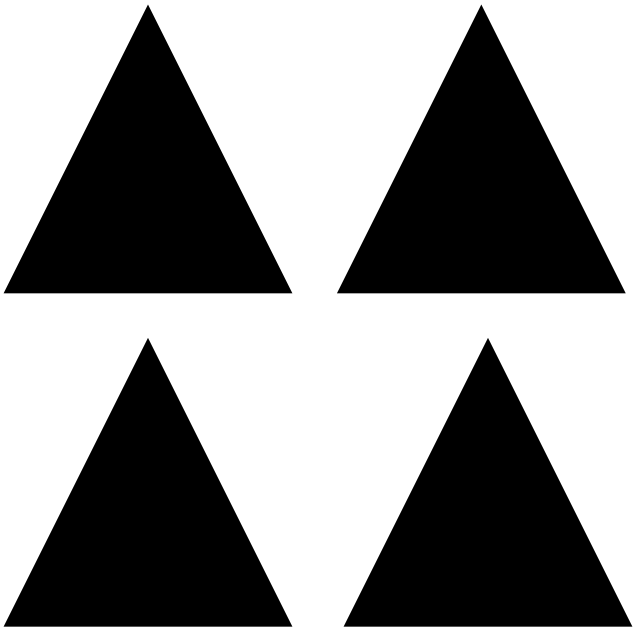}
\caption{\texttt{1000 0 d0 \\ 1020 0 d0}}
\label{fig:d0_subim4}
\end{subfigure}

\begin{subfigure}{0.45\textwidth}
\centering
\includegraphics[width=0.9\linewidth, height=4.25cm]{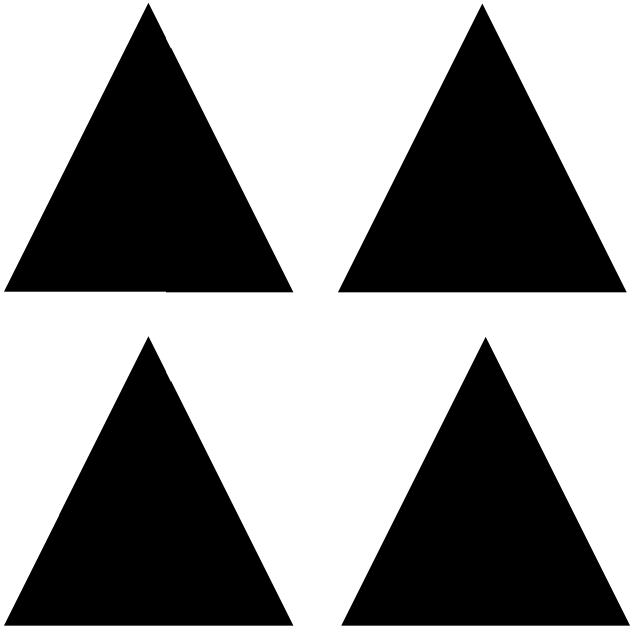}
\caption{\texttt{1000 0 d0 \\ 1010 0 d0}}
\label{fig:d0_subim5}
\end{subfigure}
\begin{subfigure}{0.45\textwidth}
\centering
\includegraphics[width=0.9\linewidth, height=4.25cm]{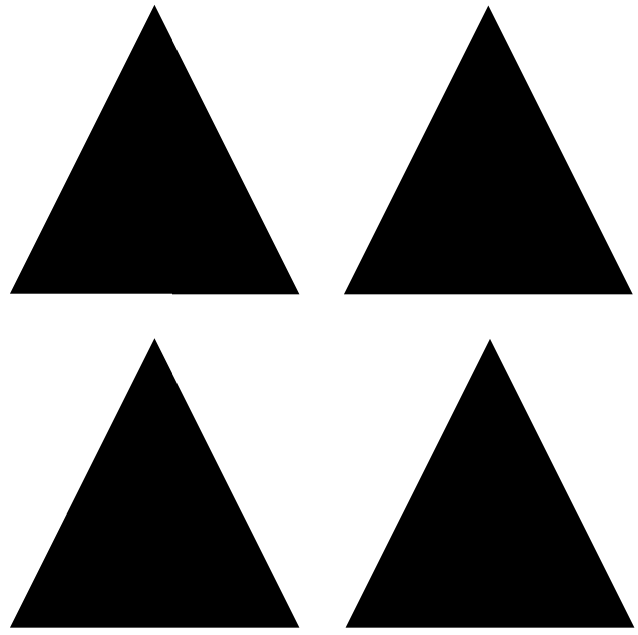}
\caption{\texttt{1000 0 d0 \\ 1005 0 d0}}
\label{fig:d0_subim6}
\end{subfigure}
\caption{d0 Operator - control,  10\%, 5\%, 2\%, 1\%, 0.5\% (1\%)}
\label{fig:d0_op}
\end{figure}

\clearpage

\subsubsection{d1 Operator}
Takes 6 floating-point operands: \texttt{$w_x \ w_y \ ll_x \ ll_y \ ur_x \ ur_y$}

Table 111: ``$w_x$ denotes the horizontal displacement in the glyph coordinate
system; it shall be consistent with the corresponding width in the
font’s Widths array. $w_y$ shall be 0 (see 9.2.4, "Glyph Positioning
and Metrics").\\
$ll_x$ and $ll_x$ denote the coordinates of the lower-left corner, and $ur_x$
and $ur_y$ denote the upper-right corner, of the glyph bounding box.
The glyph bounding box is the smallest rectangle, oriented with
the axes of the glyph coordinate system, that completely encloses
all marks placed on the page as a result of executing the glyph’s
description. The declared bounding box shall be correct—in other
words, sufficiently large to enclose the entire glyph. If any marks
fall outside this bounding box, the result is unpredictable.'' \cite{PDF_2020}

Therefore, it effectively only has 5 operands, since $w_y$ must be fixed to 0. The $w_x$ and $w_y$ operands are the same as in the \texttt{d0} operator, so they have a percentage cutoff of 1\%. The following diagrams check the percentage cutoffs for the $ll_x \ ll_y \ ur_x \ ur_y$ operands. These operands just control the bounding box around the font glyphs, which appear to only impact the box shown when the `text' is highlighted within a PDF.

\clearpage

\begin{figure}[h]
\centering
\begin{subfigure}{0.45\textwidth}
\centering
\includegraphics[width=0.4\linewidth, height=4.25cm]{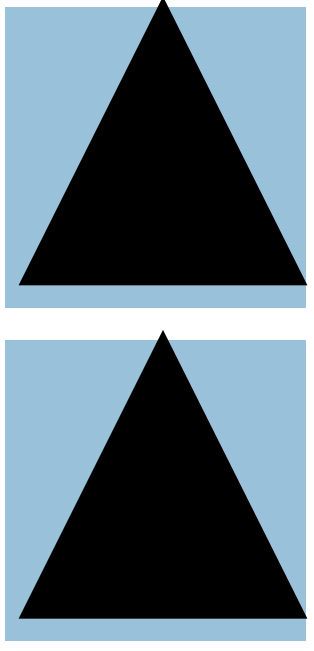}
\caption{\texttt{1000 0 -100 -100 800 800 d1 \\ 1000 0 -100 -100 800 800 d1}}
\label{fig:d1_subim1}
\end{subfigure}
\begin{subfigure}{0.45\textwidth}
\centering
\includegraphics[width=0.4\linewidth, height=4.25cm]{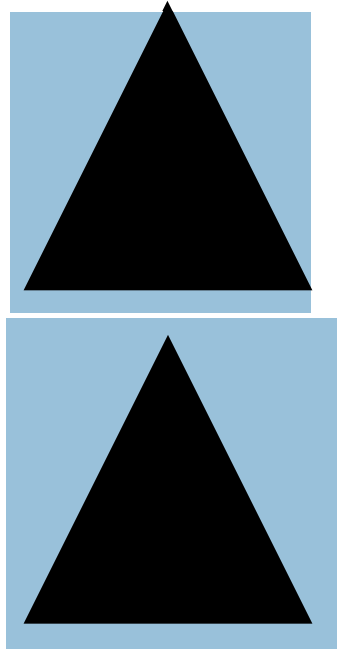}
\caption{\texttt{1000 0 -100 -100 800 800 d1 \\ 1000 0 -110 -110 880 880 d1}}
\label{fig:d1_subim2}
\end{subfigure}

\begin{subfigure}{0.45\textwidth}
\centering
\includegraphics[width=0.4\linewidth, height=4.25cm]{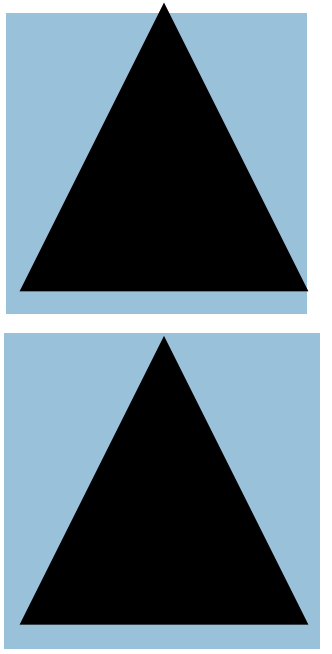}
\caption{\texttt{1000 0 -100 -100 800 800 d1 \\ 1000 0 -105 -105 840 840 d1}}
\label{fig:d1_subim3}
\end{subfigure}
\begin{subfigure}{0.45\textwidth}
\centering
\includegraphics[width=0.4\linewidth, height=4.25cm]{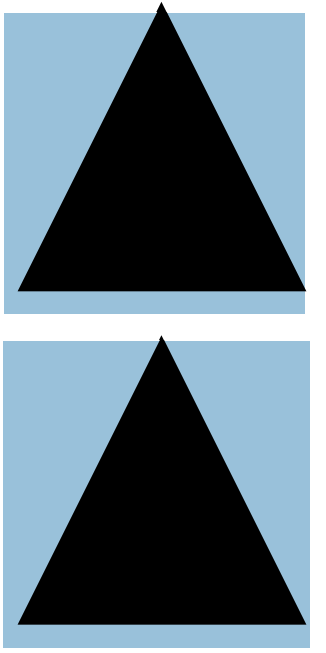}
\caption{\texttt{1000 0 -100 -100 800 800 d1 \\ 1000 0 -102 -102 816 816 d1}}
\label{fig:d1_subim4}
\end{subfigure}

\begin{subfigure}{0.45\textwidth}
\centering
\includegraphics[width=0.4\linewidth, height=4.25cm]{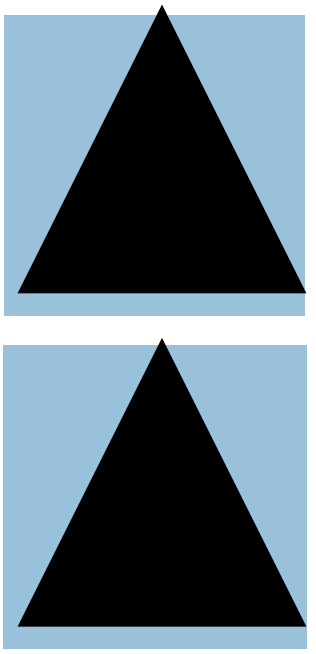}
\caption{\texttt{1000 0 -100 -100 800 800 d1 \\ 1000 0 -101 -101 808 808 d1}}
\label{fig:d1_subim5}
\end{subfigure}
\begin{subfigure}{0.45\textwidth}
\centering
\includegraphics[width=0.4\linewidth, height=4.25cm]{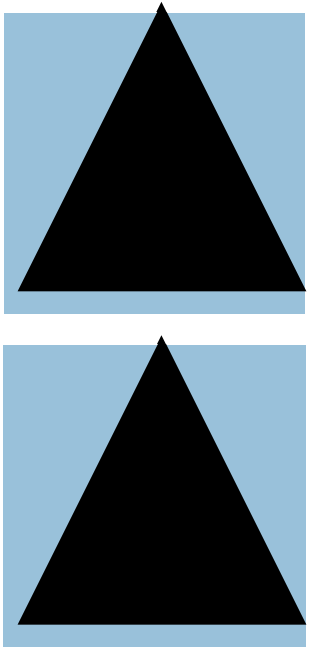}
\caption{\texttt{1000 0 -100 -100 800 800 d1 \\ 1000 0 -100.5 -100.5 804 804 d1}}
\label{fig:d1_subim6}
\end{subfigure}
\caption{d1 Operator - control,  10\%, 5\%, 2\%, 1\%, 0.5\% (1-2\%)}
\label{fig:d1_op}
\end{figure}

\end{document}